\documentclass[aps,prl,twocolumn,superscriptaddress]{revtex4}
\usepackage{amssymb,amsbsy,graphicx,xcolor,times,subfigure,marvosym,color,bm,multirow,epstopdf}
\usepackage[latin1]{inputenc}
\begin{document}

\title{Robustness of vortex-bound Majorana zero modes against
correlated disorder\footnote{This manuscript has been authored by
UT-Battelle, LLC under Contract No. DE-AC05-00OR22725 with the U.S.
Department of Energy. The United States Government retains and the
publisher, by accepting the article for publication, acknowledges
that the United States Government retains a non-exclusive, paid-up,
irrevocable, world-wide license to publish or reproduce the
published form of this manuscript, or allow others to do so, for
United States Government purposes. The Department of Energy will
provide public access to these results of federally sponsored
research in accordance with the DOE Public Access Plan
(http://energy.gov/downloads/doe-public-access-plan).}}

\author{Casey~Christian}
\affiliation{Computational Sciences and Engineering Division, Oak
Ridge National Laboratory, Oak Ridge, Tennessee 37831, USA}

\author{Eugene~F.~Dumitrescu}
\affiliation{Computational Sciences and Engineering Division, Oak
Ridge National Laboratory, Oak Ridge, Tennessee 37831, USA}

\author{G\'abor~B.~Hal\'asz}
\affiliation{Materials Science and Technology Division, Oak Ridge
National Laboratory, Oak Ridge, Tennessee 37831, USA}


\begin{abstract}

We investigate the effect of correlated disorder on Majorana zero
modes (MZMs) bound to magnetic vortices in two-dimensional
topological superconductors. By starting from a lattice model of
interacting fermions with a $p_x \pm i p_y$ superconducting ground
state in the disorder-free limit, we use perturbation theory to
describe the enhancement of the Majorana localization length at weak
disorder and a self-consistent numerical solution to understand the
breakdown of the MZMs at strong disorder. We find that correlated
disorder has a much stronger effect on the MZMs than uncorrelated
disorder and that it is most detrimental if the disorder correlation
length $\ell$ is on the same order as the superconducting coherence
length $\xi$. In contrast, MZMs can survive stronger disorder for
$\ell \ll \xi$ as random variations cancel each other within the
length scale of $\xi$, while an MZM may survive up to very strong
disorder for $\ell \gg \xi$ if it is located in a favorable domain
of the given disorder realization.

\end{abstract}


\maketitle


Topological phases of matter harbor exotic nonlocal quasiparticles
and have been proposed as a promising platform for fault-tolerant
quantum computation \cite{Kitaev-2003, Nayak-2008}. In particular,
topological superconducting systems, including one-dimensional (1D)
and two-dimensional (2D) heterostructures \cite{Fu-2008, Sau-2010,
Alicea-2010, Lutchyn-2010, Oreg-2010, Potter-2010} as well as
intrinsic 2D superconductors with $p$-wave pairing symmetry
\cite{Wang-2016}, are predicted to host Majorana zero modes (MZMs)
\cite{Kitaev-2000, Read-2000, Ivanov-2001} which can implement the
Clifford gate set via braiding. While most proposals for MZM
braiding have focused on 1D systems, such as nanowire T-junctions
\cite{Alicea-2011}, MZMs bound to superconducting vortices in 2D
systems have distinct advantages as the 2D geometry allows a greater
degree of freedom in the motion of the MZMs.

Due to their inherently nonlocal nature, MZMs are known to be
protected against infinitesimal local perturbations, including
random disorder. However, given that real-world materials contain
disorder in varying forms and strength, it is also important to
understand the robustness of MZMs against disorder beyond the
infinitesimal limit. For example, weak disorder may make the MZMs
less localized, leading to a smaller qubit density and/or more gate
errors, whereas strong disorder may lead to a complete breakdown of
the MZMs. While there have been numerous studies along these lines,
most of them focus on 1D nanowires \cite{Motrunich-2001,
Akhmerov-2011, Brouwer-2011, Lobos-2012, Bagrets-2012, Liu-2012,
Neven-2013, Sau-2013, Adagideli-2014, Hui-2014}, while those
studying 2D superconductors do not consider vortex-bound MZMs
\cite{Lu-2020} or only concentrate on uncorrelated disorder
\cite{Kraus-2009, Zhou-2017}.

In this Letter, we consider a simple microscopic model of
interacting fermions with a $p_x \pm i p_y$ superconducting ground
state \cite{Lu-1991, Cheng-2010} and study the effect of
\emph{correlated} disorder by combining analytical and numerical
approaches. Specifically, we investigate vortex-bound MZMs in this
model and understand how their robustness depends on the correlation
length of the disorder. Our main result is that correlated disorder
is significantly more detrimental to the MZMs than uncorrelated
disorder. In particular, disorder has the most adverse effect if its
correlation length $\ell$ is similar to the superconducting
coherence length $\xi$, while disorders with $\ell \ll \xi$ and
$\ell \gg \xi$ are both more benign, even though for completely
different reasons. Since our results naturally extend to the
continuum limit of the model and are expressed in terms of
measurable length and energy scales, they should apply universally
for $p_x \pm i p_y$ superconductors and provide useful guidelines
for the realization of MZM braiding in realistic experimental
systems.

\emph{Model.---}We consider a tight-binding Hamiltonian of
interacting spinless fermions on the square lattice,
\begin{eqnarray}
\hat{H} &=& -\sum_{\mathbf{r}} \mu_{\mathbf{r}}^{\phantom{\dag}}
c_{\mathbf{r}}^{\dag} c_{\mathbf{r}}^{\phantom{\dag}} -
\sum_{\langle \mathbf{r}, \mathbf{r}' \rangle} \left( t_{\mathbf{r},
\mathbf{r}'}^{\phantom{\dag}} c_{\mathbf{r}}^{\dag}
c_{\mathbf{r}'}^{\phantom{\dag}} + \mathrm{h.c.} \right)
\nonumber \\
&& - g \sum_{\langle \mathbf{r}, \mathbf{r}' \rangle}
c_{\mathbf{r}}^{\dag} c_{\mathbf{r}}^{\phantom{\dag}}
c_{\mathbf{r}'}^{\dag} c_{\mathbf{r}'}^{\phantom{\dag}},
\label{eq-H-1}
\end{eqnarray}
where the three terms describe a site-dependent chemical potential,
a nearest-neighbor hopping amplitude, and a nearest-neighbor
attractive interaction, respectively. In the presence of a magnetic
field, the hopping amplitude is spatially modulated by the vector
potential $\mathbf{A} (\mathbf{r})$ through the Peierls
substitution, $t_{\mathbf{r}, \mathbf{r}'} = t e^{i A_{\mathbf{r},
\mathbf{r}'}}$, where $A_{\mathbf{r}, \mathbf{r}'} =
\int_{\mathbf{r}}^{\mathbf{r}'} \mathbf{A} (\hat{\mathbf{r}}) \cdot
d \hat{\mathbf{r}}$. We expand the chemical potential as
$\mu_{\mathbf{r}} = \bar{\mu} + \delta \mu_{\mathbf{r}}$, where
$\bar{\mu}$ is a constant background, while $\delta
\mu_{\mathbf{r}}$ describes random disorder of strength $\delta
\bar{\mu}$ that is correlated within a length scale $\ell$.
Mathematically, $\delta \mu_{\mathbf{r}}$ are real Gaussian random
variables characterized by
\begin{equation}
\overline{\delta \mu_{\mathbf{r}}} = 0, \quad \overline{\delta
\mu_{\mathbf{r}} \delta \mu_{\mathbf{r}'}} = \delta \bar{\mu}^2
e^{-|\mathbf{r} - \mathbf{r}'|^2 / \ell^2}, \label{eq-mu-1}
\end{equation}
where the overline denotes averaging over many disorder
realizations. In practice, these real-space random variables are
generated through $\delta \mu_{\mathbf{r}} = \sum_{\mathbf{k}}
\mathrm{Re} [\delta \tilde{\mu}_{\mathbf{k}} e^{i \mathbf{k} \cdot
\mathbf{r}}]$ from the independent momentum-space complex variables
$\delta \tilde{\mu}_{\mathbf{k}}$ satisfying
\begin{eqnarray}
&& \overline{\delta \tilde{\mu}_{\mathbf{k}}^{\phantom{*}}} =
\overline{\delta \tilde{\mu}_{\mathbf{k}}^{*}} = 0, \quad
\overline{\delta \tilde{\mu}_{\mathbf{k}}^{\phantom{*}} \delta
\tilde{\mu}_{\mathbf{k}'}^{\phantom{*}}} = \overline{\delta
\tilde{\mu}_{\mathbf{k}}^{*} \delta \tilde{\mu}_{\mathbf{k}'}^{*}} =
0, \nonumber \\
&& \overline{\delta \tilde{\mu}_{\mathbf{k}}^{\phantom{*}} \delta
\tilde{\mu}_{\mathbf{k}'}^{*}} = 2 \mathcal{N} \delta \bar{\mu}^2
\delta_{\mathbf{k}, \mathbf{k}'} e^{-\ell^2 |\mathbf{k}|^2 / 4},
\label{eq-mu-2}
\end{eqnarray}
where the normalization constant is $\mathcal{N} =
[\sum_{\mathbf{k}} e^{-\ell^2 |\mathbf{k}|^2 / 4}]^{-1}$ for a large
enough system size $L \gg \ell$.

In the absence of interactions ($g = 0$), disorder
($\mu_{\mathbf{r}} = \bar{\mu}$), and magnetic field
($t_{\mathbf{r}, \mathbf{r}'} = t$), the tight-binding Hamiltonian
in Eq.~(\ref{eq-H-1}) is quadratic and translation invariant. By
means of a Fourier transform, one then obtains a single fermion band
with energy-momentum dispersion $\varepsilon_{\mathbf{k}} =
-\bar{\mu} - 2t (\cos k_x + \cos k_y)$ for a normalized lattice
constant $a = 1$. For $|\bar{\mu}| < 4t$, the low-energy physics is
governed by a Fermi surface characterized by
$\varepsilon_{\mathbf{k}} = 0$. In the following, we consider
$\bar{\mu} = -4t + \varepsilon_F$ with $0 < \varepsilon_F < t$ to
get an approximately circular Fermi surface around $\mathbf{k} =
\mathbf{0}$. From an expansion to the lowest order in $\mathbf{k}$,
the energy-momentum dispersion is then $\varepsilon_{\mathbf{k}} =
-\varepsilon_F + |\mathbf{k}|^2 / 2m$, where $m = 1 / (2t)$ is an
effective mass. Thus, in this approximation, the Fermi surface is
indeed circular with Fermi energy $\varepsilon_F$ and Fermi wave
vector $k_F = \sqrt{2m \varepsilon_F} = \sqrt{\varepsilon_F / t}$.

\emph{Bulk superconductivity.---}We first consider the Hamiltonian
in Eq.~(\ref{eq-H-1}) with attractive interactions ($g > 0$) but
without disorder ($\mu_{\mathbf{r}} = \bar{\mu}$) or magnetic field
($t_{\mathbf{r}, \mathbf{r}'} = t$). It has been shown numerically
\cite{Lu-1991} and analytically \cite{Cheng-2010} that the ground
state is then a gapped $p_x \pm i p_y$ superconductor which
spontaneously breaks time-reversal symmetry. To describe this ground
state on the mean-field (i.e., saddle-point) level, we employ a
standard Hubbard-Stratonovich decoupling in Eq.~(\ref{eq-H-1}) to
obtain a quadratic Bogoliubov-de Gennes (BdG) Hamiltonian,
\begin{eqnarray}
H &=& -\sum_{\mathbf{r}} \mu_{\mathbf{r}}^{\phantom{\dag}}
c_{\mathbf{r}}^{\dag} c_{\mathbf{r}}^{\phantom{\dag}} -
\sum_{\langle \mathbf{r}, \mathbf{r}' \rangle} \left( t_{\mathbf{r},
\mathbf{r}'}^{\phantom{*}} c_{\mathbf{r}}^{\dag}
c_{\mathbf{r}'}^{\phantom{\dag}} + t_{\mathbf{r}, \mathbf{r}'}^{*}
c_{\mathbf{r}'}^{\dag} c_{\mathbf{r}}^{\phantom{\dag}} \right)
\nonumber \\
&& - \sum_{\langle \mathbf{r}, \mathbf{r}' \rangle} \left(
\Delta_{\mathbf{r}, \mathbf{r}'}^{*} c_{\mathbf{r}}^{\phantom{\dag}}
c_{\mathbf{r}'}^{\phantom{\dag}} + \Delta_{\mathbf{r},
\mathbf{r}'}^{\phantom{*}} c_{\mathbf{r}'}^{\dag}
c_{\mathbf{r}}^{\dag} \right), \label{eq-H-2}
\end{eqnarray}
which must be solved self-consistently in terms of the
superconducting pairing potentials,
\begin{equation}
\Delta_{\mathbf{r}, \mathbf{r}'} = -\Delta_{\mathbf{r}', \mathbf{r}}
= g \langle c_{\mathbf{r}} c_{\mathbf{r}'} \rangle,
\label{eq-delta-1}
\end{equation}
where $\langle \mathcal{O} \rangle$ means the expectation value of
the operator $\mathcal{O}$ with respect to the ground state of $H$.
These pairing potentials can generally be parameterized as
\begin{eqnarray}
\Delta_{\mathbf{r}}^x \equiv \Delta_{\mathbf{r}, \mathbf{r} +
\hat{\mathbf{r}}_x} &=& \sum_{\mathbf{q}} \left(
\Delta_{\mathbf{q}}^{+} + \Delta_{\mathbf{q}}^{-} \right) e^{i
\mathbf{q} \cdot \mathbf{r}},
\nonumber \\
\Delta_{\mathbf{r}}^y \equiv \Delta_{\mathbf{r}, \mathbf{r} +
\hat{\mathbf{r}}_y} &=& i \sum_{\mathbf{q}} \left(
\Delta_{\mathbf{q}}^{+} - \Delta_{\mathbf{q}}^{-} \right) e^{i
\mathbf{q} \cdot \mathbf{r}}, \label{eq-delta-2}
\end{eqnarray}
where $\hat{\mathbf{r}}_x = (1,0)$ and $\hat{\mathbf{r}}_y = (0,1)$
are the lattice vectors, and the component
$\Delta_{\mathbf{q}}^{\pm}$ corresponds to $p_x \pm i p_y$
superconductivity with a spatial modulation of wave vector
$\mathbf{q}$. In the absence of disorder ($\mu_{\mathbf{r}} =
\bar{\mu}$) and magnetic field ($t_{\mathbf{r}, \mathbf{r}'} = t$),
the superconductivity is translation symmetric \cite{Lu-1991,
Cheng-2010}. Assuming $p_x + i p_y$ pairing symmetry without loss of
generality, the components in Eq.~(\ref{eq-delta-2}) then become
\begin{equation}
\Delta_{\mathbf{q}}^{+} = \bar{\Delta} \delta_{\mathbf{q},
\mathbf{0}}, \qquad \Delta_{\mathbf{q}}^{-} = 0, \label{eq-delta-3}
\end{equation}
corresponding to $\Delta_{\mathbf{r}} \equiv \Delta_{\mathbf{r}}^x =
-i \Delta_{\mathbf{r}}^y = \bar{\Delta}$. The constant
$\bar{\Delta}$ can be determined from a self-consistent solution of
Eqs.~(\ref{eq-H-2}) and (\ref{eq-delta-1}). In the universal
continuum limit ($k_F \ll 1$), we show in the Supplemental Material
(SM) \cite{SM} that $\bar{\Delta}$ satisfies
\begin{equation}
1 = \frac{g} {N} \sum_{\mathbf{k}} \frac{|\mathbf{k}|^2}
{\sqrt{\varepsilon_{\mathbf{k}}^2 + 4 |\mathbf{k}|^2
|\bar{\Delta}|^2}} \approx g \nu \int \frac{d\varepsilon \, k_F^2}
{\sqrt{\varepsilon^2 + 4 k_F^2 |\bar{\Delta}|^2}}, \label{eq-sc}
\end{equation}
where $N$ is the number of lattice sites, and $\nu$ is the density
of states at the Fermi level. If we then choose $\bar{\Delta}$ to be
real and positive without loss of generality, it is approximately
given by the standard superconducting gap formula,
\begin{equation}
\bar{\Delta} \sim \frac{E} {2 k_F} \exp \left( -\frac{1} {2 g k_F^2
\nu} \right), \label{eq-delta-4}
\end{equation}
where $E$ is an energy scale governing the high-energy cutoff (whose
precise value is irrelevant), while $2 g k_F^2$ is an effective
interaction strength reflecting the $p$-wave symmetry of the
superconductivity. Importantly, because of the factor $k_F^2 \propto
\varepsilon_F$ within the exponential, the pairing potential
$\bar{\Delta}$ strongly depends on the Fermi energy $\varepsilon_F$
\cite{Cheng-2010}.

Next, we include a weak disorder in the chemical potential ($\delta
\bar{\mu} \ll \bar{\mu}$) and study its effect on the pairing
potentials $\Delta_{\mathbf{r}, \mathbf{r}'}$ via perturbation
theory. Formally, we restore $\mu_{\mathbf{r}} = \bar{\mu} + \delta
\mu_{\mathbf{r}}$ in Eq.~(\ref{eq-H-2}) and modify
Eq.~(\ref{eq-delta-3}) by writing $\Delta_{\mathbf{q}}^{+} =
\bar{\Delta} \delta_{\mathbf{q}, \mathbf{0}} + \delta
\Delta_{\mathbf{q}}^{+}$ and $\Delta_{\mathbf{q}}^{-} = \delta
\Delta_{\mathbf{q}}^{-}$. We can then employ $\delta
\mu_{\mathbf{r}} = \sum_{\mathbf{q}} \delta \hat{\mu}_{\mathbf{q}}
e^{i \mathbf{q} \cdot \mathbf{r}}$, where $\delta
\hat{\mu}_{\mathbf{q}}^{\phantom{*}} = \frac{1}{2} [\delta
\tilde{\mu}_{\mathbf{q}}^{\phantom{*}} + \delta
\tilde{\mu}_{-\mathbf{q}}^{*}]$, and obtain the self-consistent
solution of Eqs.~(\ref{eq-H-2}) and (\ref{eq-delta-1}) up to linear
order in $\delta \hat{\mu}_{\mathbf{q}}$ and $\delta
\Delta_{\mathbf{q}}^{\pm}$. In the continuum limit ($|\mathbf{q}|
\ll k_F \ll 1$) of weak superconductivity ($\xi^{-1} \ll k_F$), this
approach gives (see the SM \cite{SM})
\begin{eqnarray}
\delta \Delta_{\mathbf{q}}^{+} &=& f \left( \frac{\xi |\mathbf{q}|}
{2} \right) \frac{\partial \bar{\Delta}} {\partial \varepsilon_F} \,
\delta \hat{\mu}_{\mathbf{q}},
\nonumber \\
\delta \Delta_{\mathbf{q}}^{-} &=& -h \left( \frac{\xi |\mathbf{q}|}
{2} \right) e^{2i \vartheta_{\mathbf{q}}} \, \frac{\partial
\bar{\Delta}} {\partial \varepsilon_F} \, \delta
\hat{\mu}_{\mathbf{q}}, \label{eq-delta-5}
\end{eqnarray}
where $\xi = v_F / (2k_F \bar{\Delta}) = 1 / (2m \bar{\Delta})$ is
the superconducting coherence length, $v_F = k_F / m$ is the Fermi
velocity, $\vartheta_{\mathbf{q}}$ is the angle between $\mathbf{q}$
and $\hat{\mathbf{r}}_x$, while $f(x)$ and $h(x)$ are dimensionless
functions with asymptotic forms
\begin{eqnarray}
f(x) &\approx& \bigg{\{} \begin{array}{c} 1 - \frac{x^2}{6} \quad (x
\ll 1), \\ \frac{1} {\ln x} \qquad \,\, (x \gg 1),
\end{array}
\nonumber \\
h(x) &\approx& \bigg{\{} \begin{array}{c} \frac{x^2}{6} \qquad
\,\,\,\, (x \ll 1), \\ \frac{1} {2 (\ln x)^2} \,\,\,\,\, (x \gg 1).
\end{array}
\label{eq-fh}
\end{eqnarray}
For $\mathbf{q} = \mathbf{0}$, the disorder component $\delta
\hat{\mu}_{\mathbf{0}}$ simply corresponds to a shift in the Fermi
energy $\varepsilon_F$, and the pairing potential $\bar{\Delta}$
with $p_x + i p_y$ symmetry is renormalized accordingly. For finite
$\mathbf{q}$, however, the disorder gives rise to reduced variations
in the $p_x + i p_y$ pairing due to $f (\xi |\mathbf{q}| / 2) < 1$
and also generates a finite $p_x - i p_y$ pairing due to $h (\xi
|\mathbf{q}| / 2) > 0$. Both of these effects are more pronounced if
the disorder wave vector $\mathbf{q}$ exceeds the inverse coherence
length $\xi^{-1}$. We note that, while the mean-field results in
Eqs.~(\ref{eq-delta-5}) and (\ref{eq-fh}) may not be quantitatively
right for $|\mathbf{q}| \gg \xi^{-1}$, any corrections beyond the
mean-field level are expected to strengthen our main conclusions by
suppressing $f (\xi |\mathbf{q}| / 2)$ and $h (\xi |\mathbf{q}| /
2)$.

Finally, we describe the real-space correlations in the pairing
potentials $\Delta_{\mathbf{r}, \mathbf{r}'}$ as a result of
disorder. Since $h(x) \ll f(x)$ for all $x$, we neglect the
components $\delta \Delta_{\mathbf{q}}^{-}$ and use
Eq.~(\ref{eq-delta-2}) to introduce $\delta \Delta_{\mathbf{r}}
\equiv \delta \Delta_{\mathbf{r}}^x = -i \delta
\Delta_{\mathbf{r}}^y = \sum_{\mathbf{q}} \delta
\Delta_{\mathbf{q}}^{+} e^{i \mathbf{q} \cdot \mathbf{r}}$. From
Eqs.~(\ref{eq-mu-2}) and (\ref{eq-delta-5}), the disorder
correlations in $\delta \Delta_{\mathbf{r}}$ are then
\begin{equation}
\overline{\delta \Delta_{\mathbf{r}} \delta \Delta_{\mathbf{r}'}} =
\mathcal{N} \alpha^2 \delta \bar{\mu}^2 \mathrm{Re}
\sum_{\mathbf{q}} e^{-\frac{1}{4} \ell^2 |\mathbf{q}|^2 + i
\mathbf{q} \cdot (\mathbf{r} - \mathbf{r}')} f^2 \left( \frac{\xi
|\mathbf{q}|} {2} \right), \label{eq-delta-6}
\end{equation}
where $\alpha = \partial \bar{\Delta} / \partial \varepsilon_F$ and
$f^2 (x) \equiv [f(x)]^2$. Since $f(x)$ depends only logarithmically
on its argument, it is a reasonable approximation to substitute $f
(\xi |\mathbf{q}| / 2)$ with $f (\xi / \ell)$ in
Eq.~(\ref{eq-delta-6}) and work with the resulting simplified
correlations,
\begin{equation}
\overline{\delta \Delta_{\mathbf{r}} \delta \Delta_{\mathbf{r}'}} =
\alpha^2 f^2 (\xi / \ell) \delta \bar{\mu}^2 e^{-|\mathbf{r} -
\mathbf{r}'|^2 / \ell^2}. \label{eq-delta-7}
\end{equation}
From a direct comparison with Eq.~(\ref{eq-mu-1}), this result has a
simple physical interpretation. For $\ell \gg \xi$, the local
pairing potential is determined by the local chemical potential via
Eq.~(\ref{eq-delta-4}). For $\ell \ll \xi$, the variations in the
pairing potential still follow those in the chemical potential, but
the constant of proportionality is reduced by a factor $f^2 (\xi /
\ell) \ll 1$.

\emph{Majorana localization length.---}We now consider a
superconducting vortex hosting a MZM and understand the effect of
weak disorder on the localization length of the MZM. Taking the
continuum limit, $c_{\mathbf{r}} \rightarrow \psi (\mathbf{r})$,
assuming a pure $p_x + i p_y$ pairing symmetry, $\Delta_{\mathbf{r}}
\equiv \Delta_{\mathbf{r}}^x = -i \Delta_{\mathbf{r}}^y \rightarrow
\Delta (\mathbf{r})$, and including a magnetic field, the BdG
Hamiltonian in Eq.~(\ref{eq-H-2}) takes the form $H = \int d^2
\mathbf{r} \, \bm{\psi}^{\dag} \cdot \mathcal{H} \cdot \bm{\psi}$,
where $\bm{\psi} = (\psi, \psi^{\dag})^T$ and
\begin{equation}
\mathcal{H} = \frac{1}{2} \left[
\begin{array}{cc} \frac{1}{2m} (-i \nabla + \mathbf{A})^2 - \varepsilon_F &
2 \Delta (\partial_x + i \partial_y) \\ -2 \Delta^{*} (\partial_x -
i \partial_y) & -\frac{1}{2m} (-i \nabla - \mathbf{A})^2 +
\varepsilon_F \end{array} \right]. \label{eq-H-3}
\end{equation}
Focusing on a single vortex at the origin and using polar
coordinates, $\mathbf{r} = (r, \vartheta)$, the $\pi$ magnetic flux
of the vortex can be represented by a vector potential with
components
\begin{equation}
A_r (\mathbf{r}) = 0, \qquad A_{\vartheta} (\mathbf{r}) = \frac{2\pi
\delta (\vartheta) - a(r)} {2r}, \label{eq-A}
\end{equation}
where the term $\propto \delta (\vartheta)$ corresponds to a
$\mathbb{Z}_2$ flux string \cite{Kitaev-2003, Kitaev-2006}, while
$a(r) \approx 1$ for $r \ll \lambda$ and $a(r) \sim e^{-r /
\lambda}$ for $r \gg \lambda$ in terms of the London penetration
depth $\lambda$. In this gauge, the pairing potential $\Delta
(\mathbf{r})$ does not have any angular winding and simply takes its
bulk value for $r \gg \xi$. The Hamiltonian matrix in
Eq.~(\ref{eq-H-3}) can then be written in polar coordinates as
\begin{equation}
\mathcal{H} = \frac{1}{2} \left[
\begin{array}{cc} -\frac{1}{2m} (D_r^2 + D_{\vartheta, -}^2) -
\varepsilon_F & 2 \Delta e^{i \vartheta} \big( \partial_r
+ \frac{i}{r} \partial_{\vartheta} \big) \\
-2 \Delta^{*} e^{-i \vartheta} \left( \partial_r - \frac{i}{r}
\partial_{\vartheta} \right) & \frac{1}{2m} (D_r^2 +
D_{\vartheta, +}^2) + \varepsilon_F
\end{array} \right], \label{eq-H-4}
\end{equation}
where $D_r^2 \equiv \partial_r^2 + (1/r) \partial_r$ and
$D_{\vartheta, \pm} \equiv (\partial_{\vartheta} \pm i a/2) / r$,
while the $\mathbb{Z}_2$ flux string induces antiperiodic boundary
conditions, $\psi (r, 2\pi) = -\psi (r, 0)$, in the polar angle
$\vartheta$. If we take $\Delta \in \mathbb{R}$ without loss of
generality, search for the MZM in the form
\begin{equation}
\gamma = \int d^2 \mathbf{r} \, \phi(r) \left[ i e^{-i \vartheta /
2} \, \psi (\mathbf{r}) - i e^{i \vartheta / 2} \, \psi^{\dag}
(\mathbf{r}) \right], \label{eq-gamma}
\end{equation}
which naturally satisfies the antiperiodic boundary conditions, and
demand $\gamma = \gamma^{\dag}$ as well as $[H, \gamma] = 0$, the
radial MZM wave function $\phi(r)$ must be a real solution of
\begin{equation}
\frac{1}{2m} \left[ \frac{d^2 \phi} {dr^2} + \frac{1}{r}
\frac{d\phi} {dr} - \frac{(1-a)^2 \phi} {4 r^2} \right] +
\varepsilon_F \phi + 2 \Delta \left[ \frac{d\phi} {dr} +
\frac{\phi}{2r} \right] = 0. \label{eq-phi-1}
\end{equation}
For large distances, $r \gg \lambda$, in the disorder-free limit, we
can set $\Delta = \bar{\Delta}$ and neglect $a(r) \sim e^{-r /
\lambda} \ll 1$. The \emph{exact} general solution of
Eq.~(\ref{eq-phi-1}) then takes the form
\begin{eqnarray}
\phi(r) &=& \frac{C} {\sqrt{r}} \exp \left( -2m \bar{\Delta} r
\right) \cos \left[ \sqrt{2m \varepsilon_F - (2m \bar{\Delta})^2} \,
r + \varphi \right]
\nonumber \\
&=& \frac{C} {\sqrt{r}} \exp \left( -\frac{r} {\xi} \right) \cos
\left[ q_F r + \varphi \right], \label{eq-phi-2}
\end{eqnarray}
where $C$ and $\varphi$ are arbitrary constants, while $\xi$ is the
coherence length and $q_F = \sqrt{k_F^2 - \xi^{-2}} \approx k_F$ is
the Fermi wave vector for weak superconductivity. Importantly, the
solution in Eq.~(\ref{eq-phi-2}) is approximately valid even for
$\xi \ll r \ll \lambda$ as the correction to Eq.~(\ref{eq-phi-1})
from a finite $a(r)$ is subdominant due to $|\phi / r^2| \ll |d^2
\phi / dr^2|$ for any $r \gg \xi$. As expected, the Majorana
localization length is thus simply the coherence length $\xi$ in the
disorder-free limit.

If we include a weak disorder in the chemical potential $\mu$ (i.e.,
the Fermi energy $\varepsilon_F$), it affects the decay of the MZM
wave function $\phi(r)$ and, hence, the localization length via the
pairing potential $\Delta$. Ignoring the power-law prefactor, the
approximate disorder average of $|\phi(r)|$ from
Eq.~(\ref{eq-phi-2}) is
\begin{equation}
\overline{|\phi(r)|} \sim \overline{\exp \left( -2m \int_0^r d
\hat{r} \left[ \bar{\Delta} + \delta \Delta (\hat{r}) \right]
\right)}. \label{eq-phi-3}
\end{equation}
Utilizing the Gaussian nature of the random component $\delta \Delta
(\hat{r})$ and taking its correlations from Eq.~(\ref{eq-delta-7}),
the disorder average for $r \gg \ell \gg k_F^{-1}$ then becomes
\begin{eqnarray}
\overline{|\phi(r)|} &\sim& \exp \left[ -2m \bar{\Delta} r + 2m^2
\int_0^r d \hat{r} \int_0^r d \hat{r}' \, \overline{\delta \Delta
(\hat{r}) \delta \Delta (\hat{r}')} \right] \nonumber \\
&\approx& \exp \left[ -\frac{r} {\xi} + \frac{\sqrt{\pi}} {2}
\kappa^2 \frac{\ell r} {\xi^2} \, f^2 \left( \frac{\xi} {\ell}
\right) \right], \label{eq-phi-4}
\end{eqnarray}
and corresponds to an enhanced Majorana localization length
\begin{equation}
\xi' = \xi \left[ 1 - \frac{\sqrt{\pi}} {2} \kappa^2 \frac{\ell}
{\xi} \, f^2 \left( \frac{\xi} {\ell} \right) \right]^{-1},
\label{eq-xi}
\end{equation}
where $\kappa = \delta \bar{\mu} (\partial \bar{\Delta} / \partial
\varepsilon_F) / \bar{\Delta}$ is the relative change in the pairing
potential as a result of a shift $\delta \bar{\mu}$ in the Fermi
energy. According to Eq.~(\ref{eq-xi}), the localization length is
more sensitive to disorder with larger correlation length $\ell$.
Indeed, for $\ell \ll \xi$, the correction to the localization
length is suppressed due to both $\ell / \xi \ll 1$ and $f (\xi /
\ell) \ll 1$. However, for $\ell \gg \xi$, it should be emphasized
that the disorder average leading to Eq.~(\ref{eq-xi}) is only
appropriate for $r \gg \ell$. Instead, for $\xi \ll r \ll \ell$, the
behavior is determined by the specific disorder realization, and the
localization length may even \emph{decrease} if the MZM is located
in a region with $\mu > \bar{\mu}$.

\begin{figure}[t]
\centering
\includegraphics[width=0.99\columnwidth]{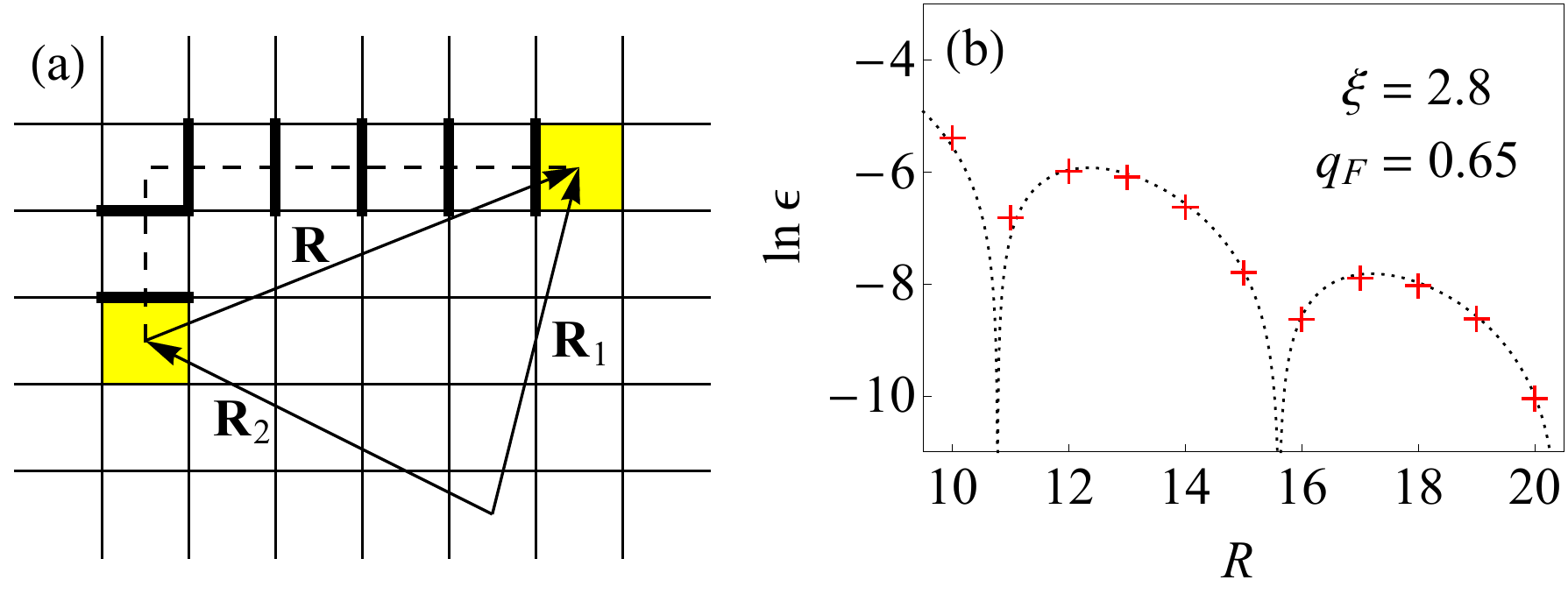}
\caption{(a) Two vortices centered at the yellow plaquettes with
separation $\mathbf{R} = \mathbf{R}_1 - \mathbf{R}_2 = (5,2)$. The
$\mathbb{Z}_2$ flux string (dashed line) intersects several links
denoted by thick lines. (b) MZM hybridization energy $\epsilon$ as a
function of the separation $\mathbf{R} = (R,0)$ for a $50 \times 30$
system. The dotted line is a fit of Eq.~(\ref{eq-phi-2}) with $\xi =
2.8$ and $q_F = 0.65$.} \label{fig-1}
\end{figure}

\emph{Numerical solution.---}To qualitatively check the validity of
our results, we numerically obtain self-consistent solutions of
Eqs.~(\ref{eq-H-2}) and (\ref{eq-delta-1}) through an iterative
procedure. Since MZMs must appear in pairs for any closed system, we
consider two superconducting vortices centered at two square
plaquettes with positions $\mathbf{R}_{1,2}$ [see
Fig.~\ref{fig-1}(a)]. In this case, the $\mathbb{Z}_2$ flux string
connects the two vortices, and the hopping amplitudes in
Eq.~(\ref{eq-H-2}) become $t_{\mathbf{r}, \mathbf{r}'} = t
u_{\mathbf{r}, \mathbf{r'}} e^{i A'_{\mathbf{r}, \mathbf{r}'}}$,
where $u_{\mathbf{r}, \mathbf{r'}}$ is $-1$ ($+1$) if the
$\mathbb{Z}_2$ flux string intersects (does not intersect) the link
$\langle \mathbf{r}, \mathbf{r}' \rangle$, while $A'_{\mathbf{r},
\mathbf{r}'}$ is only nonzero within a radius $\lambda$ of each
vortex. The precise form of $A'_{\mathbf{r}, \mathbf{r}'}$ and the
details of the iterative procedure are described in the SM \cite{SM,
Footnote}.

We choose the parameters of Eq.~(\ref{eq-H-1}) to be $t = 1$,
$\bar{\mu} = -3.5$, and $g = 5.0$, which correspond to $m = 0.5$,
$\varepsilon_F = 0.5$, and $k_F \approx 0.7$. In the absence of
disorder, the self-consistent solution for a vortex-free system
gives a bulk pairing potential $\bar{\Delta} \approx 0.33$ and a
bulk fermion gap $E_0 \approx 0.41$. If we then include two vortices
with separation $\mathbf{R} = \mathbf{R}_1 - \mathbf{R}_2$, we find
a low-energy fermion in the bulk gap whose energy decays
exponentially with $R \equiv |\mathbf{R}|$ [see
Fig.~\ref{fig-1}(b)]. Since this fermion consists of the two MZMs
bound to the vortices, and its finite energy results from a
hybridization between the MZM wave functions, we fit its energy
$\epsilon$ with the functional form of Eq.~(\ref{eq-phi-2}) to
extract $\xi \approx 2.8$ and $q_F \approx 0.65$. We note that these
values agree with $1 / (2m \bar{\Delta}) \approx 3.0$ and $k_F
\approx 0.7$ even though the system is not in the continuum limit.

\begin{figure}[t]
\centering
\includegraphics[width=0.99\columnwidth]{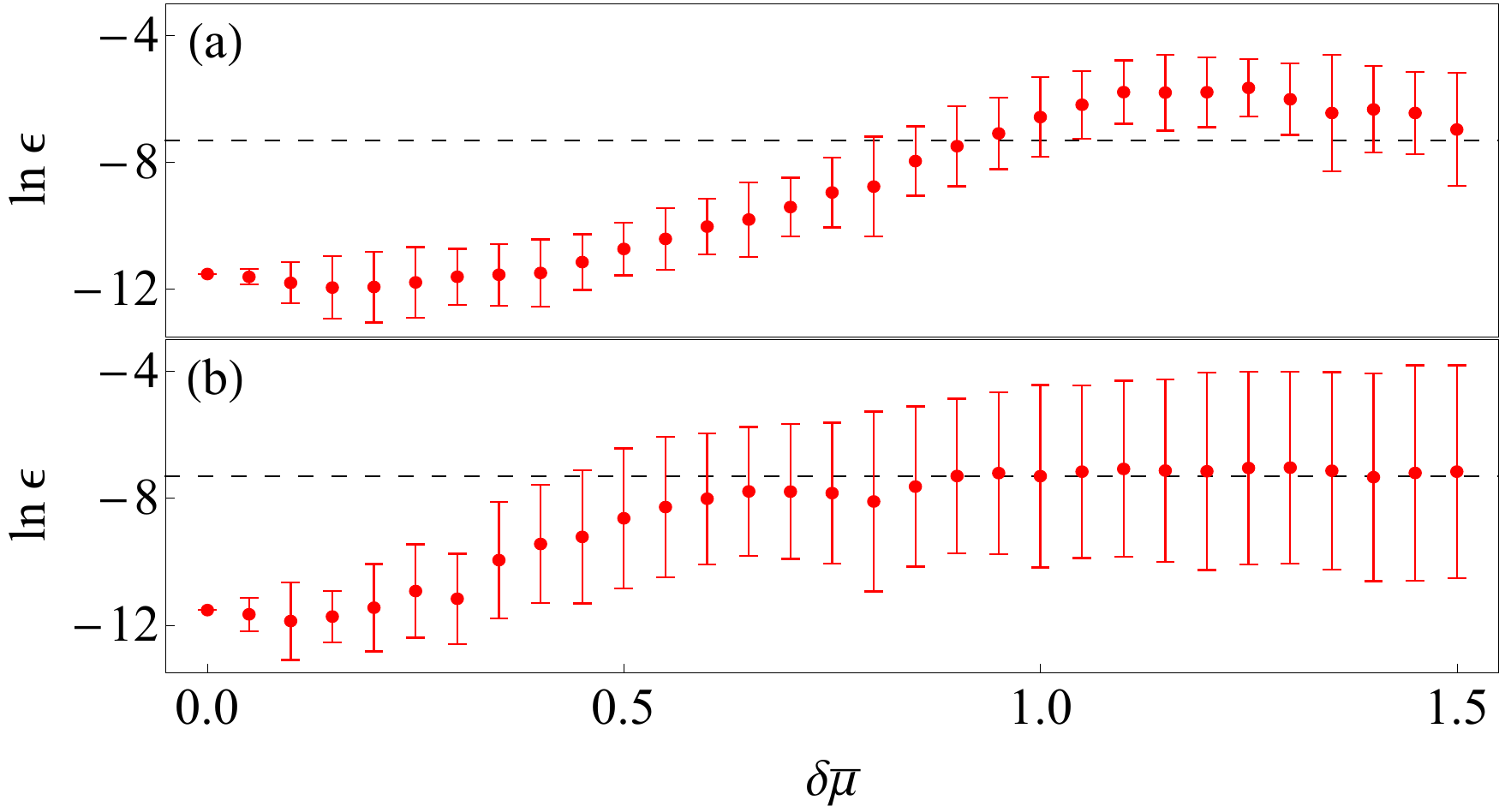}
\caption{Disorder-averaged MZM hybridization energy $\epsilon$
against disorder strength $\delta \bar{\mu}$ for (a) $\ell = 1$ and
(b) $\ell = 10$ with a separation $\mathbf{R} = (25,15)$ for a $50
\times 30$ system. Each data point is averaged over $25$ disorder
realizations, and its error bar shows the variation among the
individual realizations. The dashed line marks the expectation for a
generic disordered system, $\epsilon \sim 1/N = 1/1500$.}
\label{fig-2}
\end{figure}

Finally, we include two vortices with $R \gg 1$ and investigate how
the energy $\epsilon$ of the lowest-energy fermion behaves as the
disorder strength $\delta \bar{\mu}$ is gradually increased. The
disorder-averaged results are shown in Fig.~\ref{fig-2} for two
different disorder correlation lengths, corresponding to $\ell <
\xi$ and $\ell > \xi$, respectively. In both cases, we find that the
energy $\epsilon$ increases from the MZM result, $\epsilon \sim
e^{-R / \xi}$, to the generic disordered result, $\epsilon \sim
1/N$, which indicates the breakdown of the MZMs. This breakdown
occurs at $\delta \bar{\mu} \sim \varepsilon_F$ due to a
hybridization between the MZMs and the gapless edge modes that
surround disorder-induced non-superconducting regions with local
$\varepsilon_F < 0$. Remarkably, this breakdown is in qualitative
agreement with our weak-disorder results in at least three different
ways. First, the breakdown at $\delta \bar{\mu} \sim \varepsilon_F$
roughly corresponds to $\kappa \sim 1$ at which Eq.~(\ref{eq-xi})
predicts a divergent localization length in the case of $\ell \sim
\xi$. Second, the MZMs can \emph{generally} survive stronger
disorder for $\ell < \xi$. Third, the energy $\epsilon$ has larger
variations for $\ell > \xi$ as the MZMs can survive even very strong
disorder for \emph{certain} disorder realizations.

\emph{Discussion.---}We have studied the effect of correlated
disorder on vortex-bound MZMs in $p_x \pm i p_y$ superconductors and
demonstrated that it is much more detrimental than uncorrelated
disorder. The general picture is that disorder gradually increases
the MZM localization length until the MZMs eventually break down due
to a divergent localization length. However, according to
Eq.~(\ref{eq-xi}), the correction to the localization length
strongly depends on the disorder correlation length $\ell$ and is
suppressed for short-range-correlated disorder ($\ell \ll \xi$)
because random variations cancel each other within the
superconducting coherence length $\xi$. We note that, while
Eq.~(\ref{eq-xi}) is only valid for $\ell \gg k_F^{-1}$, our
numerical results confirm this suppression even in the uncorrelated
limit ($\ell \rightarrow 0$).

For long-range-correlated disorder ($\ell \gg \xi$), the MZM
localization length, which characterizes the decay of the wave
function $\phi(r)$ at large distances, $r \gg \ell$, is strongly
renormalized and thus rapidly diverges. Nevertheless, if the MZM is
located within a large (size $\ell$) ``disorder domain'' with $\mu
> \bar{\mu}$, it survives even in the presence of strong disorder
because its wave function is already exponentially small,
$\phi(\ell) \sim e^{-\ell / \xi}$, at the boundary, $r \sim \ell$,
of the disorder domain. While any actual braiding of the MZMs is
then restricted to such favorable disorder domains, effective
braiding may still be achievable through a measurement-only protocol
\cite{Bonderson-2008, Vijay-2016}.

Therefore, we conclude that disorder has the most adverse effect on
the MZMs if its correlation length is similar to the superconducting
coherence length. In this regime, the MZMs break down if disorder is
strong enough to induce topologically distinct regions surrounded by
gapless edge modes. We emphasize that, while we focus on a specific
lattice model and only include disorder in the chemical potential,
our results naturally extend to the continuum limit and should be
universally applicable to disordered $p_x \pm i p_y$
superconductors.

We thank Cristian Batista for useful discussions. This research was
sponsored by the U.~S.~Department of Energy, Office of Science,
Basic Energy Sciences, Materials Sciences and Engineering Division.
Preliminary modeling by G.~B.~H.~was supported by the Laboratory
Directed Research and Development Program of Oak Ridge National
Laboratory, managed by UT-Battelle, LLC, for the U.~S.~Department of
Energy. C.~C.~was partially supported by the DOE Science
Undergraduate Laboratory Internships (SULI) program.



\clearpage

\begin{widetext}

\subsection{\large Supplemental Material}

\section{Mean-field theory of bulk superconductivity}

\subsection{General formulation}

Here we derive the mean-field theory for the $p_x \pm i p_y$
superconducting ground state of our model. Employing the
path-integral formulation, the partition function corresponding to
the Hamiltonian in Eq.~(1) of the main text reads
\begin{eqnarray}
&& Z = \int D (\psi, \psi^{*}) \exp \left( -S [\psi]
\right), \label{eq-gen-Z-1} \\
&& S [\psi] = \int_0^{\beta} d\tau \, \Bigg\{ \sum_{\mathbf{r}}
\psi_{\mathbf{r}}^{*} (\tau) \left[
\partial_{\tau}^{\phantom{*}} - \mu_{\mathbf{r}}^{\phantom{*}} \right]
\psi_{\mathbf{r}}^{\phantom{*}} (\tau) - \sum_{\langle \mathbf{r},
\mathbf{r}' \rangle} \left[ t_{\mathbf{r},
\mathbf{r}'}^{\phantom{*}} \psi_{\mathbf{r}}^{*} (\tau)
\psi_{\mathbf{r}'}^{\phantom{*}} (\tau) + \mathrm{c.c.} \right] - g
\sum_{\langle \mathbf{r}, \mathbf{r}' \rangle} \psi_{\mathbf{r}}^{*}
(\tau) \psi_{\mathbf{r}}^{\phantom{*}} (\tau) \psi_{\mathbf{r}'}^{*}
(\tau) \psi_{\mathbf{r}'}^{\phantom{*}} (\tau) \Bigg\}, \nonumber
\end{eqnarray}
where $\beta$ is the inverse temperature, while
$\psi_{\mathbf{r}}^{\phantom{*}} (\tau)$ and $\psi_{\mathbf{r}}^{*}
(\tau)$ are Grassmann fields representing the fermionic operators
$c_{\mathbf{r}}^{\phantom{\dag}}$ and $c_{\mathbf{r}}^{\dag}$,
respectively. Introducing the bosonic Hubbard-Stratonovich fields
$\Delta_{\mathbf{r}, \mathbf{r}'}^{\phantom{*}} (\tau)$ and
$\Delta_{\mathbf{r}, \mathbf{r}'}^{*} (\tau)$, the partition
function then becomes
\begin{eqnarray}
&& Z = \int D (\Delta, \Delta^{*}) \int D (\psi, \psi^{*}) \exp
\left( -S [\Delta, \psi] \right), \label{eq-gen-Z-2} \\
&& S [\Delta, \psi] = \int_0^{\beta} d\tau \, \Bigg\{
\sum_{\mathbf{r}} \psi_{\mathbf{r}}^{*} (\tau) \left[
\partial_{\tau}^{\phantom{*}} - \mu_{\mathbf{r}}^{\phantom{*}} \right]
\psi_{\mathbf{r}}^{\phantom{*}} (\tau) - \sum_{\langle \mathbf{r},
\mathbf{r}' \rangle} \left[ t_{\mathbf{r},
\mathbf{r}'}^{\phantom{*}} \psi_{\mathbf{r}}^{*} (\tau)
\psi_{\mathbf{r}'}^{\phantom{*}} (\tau) + t_{\mathbf{r},
\mathbf{r}'}^{*} \psi_{\mathbf{r}'}^{*} (\tau)
\psi_{\mathbf{r}}^{\phantom{*}} (\tau) \right] \nonumber \\
&& \qquad \qquad \,\,\, -\sum_{\langle \mathbf{r}, \mathbf{r}'
\rangle} \left[ \Delta_{\mathbf{r}, \mathbf{r}'}^{*} (\tau)
\psi_{\mathbf{r}}^{\phantom{*}} (\tau)
\psi_{\mathbf{r}'}^{\phantom{*}} (\tau) + \Delta_{\mathbf{r},
\mathbf{r}'}^{\phantom{*}} (\tau) \psi_{\mathbf{r}'}^{*} (\tau)
\psi_{\mathbf{r}}^{*} (\tau) \right] + \frac{1}{g} \sum_{\langle
\mathbf{r}, \mathbf{r}' \rangle} \left| \Delta_{\mathbf{r},
\mathbf{r}'}^{\phantom{*}} (\tau) \right|^2 \Bigg\}. \nonumber
\end{eqnarray}
Since the action $S [\Delta, \psi]$ is quadratic in the Grassmann
fields $\psi_{\mathbf{r}}^{\phantom{*}} (\tau)$ and
$\psi_{\mathbf{r}}^{*} (\tau)$, these Grassmann fields can be
integrated out to obtain an effective action $S [\Delta]$
exclusively in terms of the Hubbard-Stratonovich fields
$\Delta_{\mathbf{r}, \mathbf{r}'}^{\phantom{*}} (\tau)$ and
$\Delta_{\mathbf{r}, \mathbf{r}'}^{*} (\tau)$. To this end, it is
useful to introduce Fourier transforms in both space and (imaginary)
time for both the Grassmann fields,
\begin{equation}
\psi_{\mathbf{r}}^{\phantom{*}} (\tau) = \frac{1} {\sqrt{\beta N}}
\sum_{\mathbf{k}, \omega_n} \psi_{\mathbf{k},
\omega_n}^{\phantom{*}} e^{i \mathbf{k} \cdot \mathbf{r} - i
\omega_n \tau}, \label{eq-gen-psi}
\end{equation}
as well as the Hubbard-Stratonovich fields,
\begin{eqnarray}
\Delta_{\mathbf{r}}^x (\tau) &\equiv& \Delta_{\mathbf{r}, \mathbf{r}
+ \hat{\mathbf{r}}_x}^{\phantom{*}} (\tau) = \sum_{\mathbf{q},
\Omega_n} \left( \Delta_{\mathbf{q}, \Omega_n}^{+} +
\Delta_{\mathbf{q}, \Omega_n}^{-} \right) e^{i \mathbf{q} \cdot
\mathbf{r} - i \Omega_n \tau}, \nonumber \\
\Delta_{\mathbf{r}}^y (\tau) &\equiv& \Delta_{\mathbf{r}, \mathbf{r}
+ \hat{\mathbf{r}}_y}^{\phantom{*}} (\tau) = i \sum_{\mathbf{q},
\Omega_n} \left( \Delta_{\mathbf{q}, \Omega_n}^{+} -
\Delta_{\mathbf{q}, \Omega_n}^{-} \right) e^{i \mathbf{q} \cdot
\mathbf{r} - i \Omega_n \tau}, \label{eq-gen-delta}
\end{eqnarray}
where $N$ is the number of lattice sites, $\hat{\mathbf{r}}_x =
(1,0)$ and $\hat{\mathbf{r}}_y = (0,1)$ are the lattice vectors,
$\omega_n = (2n+1) \pi / \beta$ are the fermionic Matsubara
frequencies, $\Omega_n = 2n \pi / \beta$ are the bosonic Matsubara
frequencies, while $\Delta_{\mathbf{q}, \Omega_n}^{\pm}$ correspond
to pairing potentials with $p_x \pm i p_y$ pairing symmetry and a
spatial modulation of wave vector $\mathbf{q}$ [see also Eq.~(6) in
the main text]. Setting $t_{\mathbf{r}, \mathbf{r}'} = t$ and
$\mu_{\mathbf{r}} = \bar{\mu} + \sum_{\mathbf{q}} \delta
\hat{\mu}_{\mathbf{q}} e^{i \mathbf{q} \cdot \mathbf{r}}$, the
action $S [\Delta, \psi]$ in Eq.~(\ref{eq-gen-Z-2}) can then be
written as
\begin{eqnarray}
&& S [\Delta, \psi] = \frac{2N \beta} {g} \sum_{\mathbf{q},
\Omega_n} \Big( \big| \Delta_{\mathbf{q}, \Omega_n}^{+} \big|^2 +
\big| \Delta_{\mathbf{q}, \Omega_n}^{-} \big|^2 \Big) + \frac{1}{2}
\sum_{\mathbf{k}, \omega_n} \sum_{\mathbf{k}', \omega_n'} \left(
\begin{array}{cc} \psi_{\mathbf{k}, \omega_n}^{*} & \psi_{-\mathbf{k},
-\omega_n}^{\phantom{*}} \end{array} \right) \cdot G_{(\mathbf{k},
\omega_n), (\mathbf{k}', \omega_n')}^{-1} [\Delta] \cdot \left(
\begin{array}{c} \psi_{\mathbf{k}', \omega_n'}^{\phantom{*}} \\
\psi_{-\mathbf{k}', -\omega_n'}^{*} \end{array} \right), \nonumber
\\
\label{eq-gen-S}
\\
&& G_{(\mathbf{k}, \omega_n), (\mathbf{k}', \omega_n')}^{-1}
[\Delta] = \left( \begin{array}{cc} \delta_{\omega_n, \omega_n'}
\left[ (-i \omega_n^{\phantom{x}} +
\varepsilon_{\mathbf{k}}^{\phantom{*}}) \delta_{\mathbf{k},
\mathbf{k}'}^{\phantom{*}} - \delta \hat{\mu}_{\mathbf{k} -
\mathbf{k}'}^{\phantom{*}} \right] & i p_{\mathbf{k},
\mathbf{k}'}^{\phantom{*}} \Delta_{\mathbf{k} - \mathbf{k}',
\omega_n - \omega_n'}^{+} + i p_{\mathbf{k}', \mathbf{k}}^{*}
\Delta_{\mathbf{k} - \mathbf{k}', \omega_n - \omega_n'}^{-} \\ -i
p_{\mathbf{k}', \mathbf{k}}^{*} \big( \Delta_{\mathbf{k}' -
\mathbf{k}, \omega_n' - \omega_n}^{+} \big)^{*} - i p_{\mathbf{k},
\mathbf{k}'}^{\phantom{*}} \big( \Delta_{\mathbf{k}' - \mathbf{k},
\omega_n' - \omega_n}^{-} \big)^{*} & \delta_{\omega_n, \omega_n'}
\left[ (-i \omega_n^{\phantom{x}} -
\varepsilon_{\mathbf{k}}^{\phantom{*}}) \delta_{\mathbf{k},
\mathbf{k}'}^{\phantom{*}} + \delta \hat{\mu}_{\mathbf{k} -
\mathbf{k}'}^{\phantom{*}} \right]
\end{array} \right). \nonumber
\end{eqnarray}
In the continuum limit, corresponding to $|\mathbf{k}| \sim
|\mathbf{k}'| \sim k_F \ll 1$, the functions
$\varepsilon_{\mathbf{k}}$ and $p_{\mathbf{k}, \mathbf{k}'}$ can be
expanded as
\begin{eqnarray}
\varepsilon_{\mathbf{k}} &=& -\bar{\mu} - 2t \left( \cos k_x + \cos
k_y \right) \approx -\varepsilon_F + \frac{|\mathbf{k}|^2} {2m},
\nonumber \\
p_{\mathbf{k}, \mathbf{k}'} &=& -i \left[ e^{i \mathbf{k}' \cdot
\hat{\mathbf{r}}_x} + i e^{i \mathbf{k}' \cdot \hat{\mathbf{r}}_y} -
e^{-i \mathbf{k} \cdot \hat{\mathbf{r}}_x} - i e^{-i \mathbf{k}
\cdot \hat{\mathbf{r}}_y} \right] \approx \big( k_x + i k_y \big) +
\big( k_x' + i k_y' \big), \label{eq-gen-ep}
\end{eqnarray}
where $m = 1 / (2t)$ is the effective mass, $\varepsilon_F =
\bar{\mu} + 4t$ is the Fermi energy, and $k_F = \sqrt{2m
\varepsilon_F} = \sqrt{\varepsilon_F / t}$ is the Fermi wave vector.
Integrating out the Grassmann fields, the partition function in
Eq.~(\ref{eq-gen-Z-2}) then takes the form
\begin{eqnarray}
&& Z = \int D (\Delta, \Delta^{*}) \exp
\left( -S [\Delta] \right), \label{eq-gen-Z-3} \\
&& S [\Delta] = \frac{2N \beta} {g} \sum_{\mathbf{q}, \Omega_n}
\Big( \big| \Delta_{\mathbf{q}, \Omega_n}^{+} \big|^2 + \big|
\Delta_{\mathbf{q}, \Omega_n}^{-} \big|^2 \Big) - \mathrm{Tr} \ln
G^{-1} [\Delta]. \nonumber
\end{eqnarray}
The infinitely large matrix $G^{-1} [\Delta]$ simultaneously acts in
particle-hole (Nambu) space, momentum space, and frequency space,
while its $2 \times 2$ blocks corresponding to Nambu space are given
by $G_{(\mathbf{k}, \omega_n), (\mathbf{k}', \omega_n')}^{-1}$ in
Eq.~(\ref{eq-gen-S}).

\subsection{Mean-field theory in the disorder-free limit}

On the level of mean-field theory, we restrict our attention to the
saddle points of the effective action in Eq.~(\ref{eq-gen-Z-3}).
Differentiating $S [\Delta]$ with respect to $(\Delta_{\mathbf{q},
\Omega_n}^{\pm})^{*}$, the general saddle-point equation becomes
\begin{equation}
\frac{\partial S [\Delta]} {\partial (\Delta_{\mathbf{q},
\Omega_n}^{\pm})^{*}} = \frac{2N \beta} {g} \Delta_{\mathbf{q},
\Omega_n}^{\pm} - \mathrm{Tr} \left\{ G [\Delta] \cdot
\frac{\partial G^{-1} [\Delta]} {\partial (\Delta_{\mathbf{q},
\Omega_n}^{\pm})^{*}} \right\} = 0, \label{eq-mf-S-1}
\end{equation}
where $G [\Delta]$ is the inverse matrix of $G^{-1} [\Delta]$. In
the disorder-free limit ($\delta \hat{\mu}_{\mathbf{q}} = 0$), the
saddle point with spatially homogeneous $p_x + i p_y$
superconductivity, corresponding to the known ground state of the
model, is characterized by
\begin{equation}
\Delta_{\mathbf{q}, \Omega_n}^{+} = \bar{\Delta} \,
\delta_{\mathbf{q}, \mathbf{0}} \, \delta_{\Omega_n, 0}, \qquad
\Delta_{\mathbf{q}, \Omega_n}^{-} = 0. \label{eq-mf-delta}
\end{equation}
The matrices $G^{-1} [\Delta]$ and $G [\Delta]$ are then block
diagonal in both $\mathbf{k}$ and $\omega_n$, and their respective
$2 \times 2$ blocks are given by
\begin{eqnarray}
G_{\mathbf{k}, \omega_n}^{-1} [\Delta] &\equiv& G_{(\mathbf{k},
\omega_n), (\mathbf{k}, \omega_n)}^{-1} [\Delta] = \left(
\begin{array}{cc} -i \omega_n^{\phantom{x}} +
\varepsilon_{\mathbf{k}}^{\phantom{*}} & i
p_{\mathbf{k}}^{\phantom{*}} \bar{\Delta} \\ -i p_{\mathbf{k}}^{*}
\bar{\Delta}^{*} & -i \omega_n^{\phantom{x}} -
\varepsilon_{\mathbf{k}}^{\phantom{*}}
\end{array} \right), \nonumber
\\
G_{\mathbf{k}, \omega_n} [\Delta] &\equiv& G_{(\mathbf{k},
\omega_n), (\mathbf{k}, \omega_n)} [\Delta] = \frac{1} {\omega_n^2 +
\varepsilon_{\mathbf{k}}^2 + |p_{\mathbf{k}}^{\phantom{*}}|^2
|\bar{\Delta}|^2} \left( \begin{array}{cc} i \omega_n^{\phantom{x}}
+ \varepsilon_{\mathbf{k}}^{\phantom{*}} & i
p_{\mathbf{k}}^{\phantom{*}} \bar{\Delta} \\ -i p_{\mathbf{k}}^{*}
\bar{\Delta}^{*} & i \omega_n^{\phantom{x}} -
\varepsilon_{\mathbf{k}}^{\phantom{*}}
\end{array} \right), \label{eq-mf-G}
\end{eqnarray}
where $p_{\mathbf{k}} \equiv p_{\mathbf{k}, \mathbf{k}} \approx 2
(k_x + i k_y),$ while the saddle-point equation in
Eq.~(\ref{eq-mf-S-1}) takes the form
\begin{equation}
\bar{\Delta} = \frac{g} {2N \beta} \sum_{\mathbf{k}, \omega_n}
\mathrm{Tr} \left\{ G_{\mathbf{k}, \omega_n} [\Delta] \cdot
\frac{\partial G_{\mathbf{k}, \omega_n}^{-1} [\Delta]} {\partial
\bar{\Delta}^{*}} \right\} = \frac{g} {2N \beta} \sum_{\mathbf{k},
\omega_n} \frac{|p_{\mathbf{k}}^{\phantom{*}}|^2 \bar{\Delta}}
{\omega_n^2 + \varepsilon_{\mathbf{k}}^2 +
|p_{\mathbf{k}}^{\phantom{*}}|^2 |\bar{\Delta}|^2}.
\label{eq-mf-S-2}
\end{equation}
At zero temperature ($\beta \rightarrow \infty$), the summation in
the Matsubara frequency $\omega_n$ can be turned into an integral.
Dividing both sides of Eq.~(\ref{eq-mf-S-2}) by $\bar{\Delta}$, and
using $|p_{\mathbf{k}}|^2 = 4 |\mathbf{k}|^2$, the saddle-point
equation then becomes
\begin{equation}
1 = \frac{g} {4 \pi N} \sum_{\mathbf{k}} \int_{-\infty}^{+\infty}
d\omega \, \frac{|p_{\mathbf{k}}^{\phantom{*}}|^2} {\omega^2 +
\varepsilon_{\mathbf{k}}^2 + |p_{\mathbf{k}}^{\phantom{*}}|^2
|\bar{\Delta}|^2} = \frac{g} {4N} \sum_{\mathbf{k}}
\frac{|p_{\mathbf{k}}^{\phantom{*}}|^2}
{\sqrt{\varepsilon_{\mathbf{k}}^2 + |p_{\mathbf{k}}^{\phantom{*}}|^2
|\bar{\Delta}|^2}} = \frac{g} {N} \sum_{\mathbf{k}}
\frac{|\mathbf{k}|^2} {\sqrt{\varepsilon_{\mathbf{k}}^2 + 4
|\mathbf{k}|^2 |\bar{\Delta}|^2}}. \label{eq-mf-S-3}
\end{equation}
This final form of the saddle-point equation is equivalent to
Eq.~(8) in the main text.

\subsection{Disorder corrections through perturbation theory}

In the presence of disorder ($\delta \hat{\mu}_{\mathbf{q}} \neq
0$), we consider perturbative corrections to the disorder-free
saddle point. To this end, we modify Eq.~(\ref{eq-mf-delta}) by
including spatially inhomogeneous corrections as
\begin{equation}
\Delta_{\mathbf{q}, \Omega_n}^{+} = \left( \bar{\Delta} \,
\delta_{\mathbf{q}, \mathbf{0}} + \delta \Delta_{\mathbf{q}}^{+}
\right) \delta_{\Omega_n, 0}, \qquad \Delta_{\mathbf{q},
\Omega_n}^{-} = \delta \Delta_{\mathbf{q}}^{-} \, \delta_{\Omega_n,
0}. \label{eq-corr-delta-1}
\end{equation}
The spatial inhomogeneities $\delta \hat{\mu}_{\mathbf{q}}$ and
$\delta \Delta_{\mathbf{q}}^{\pm}$ give corrections to the matrices
$G^{-1} [\Delta]$ and $G [\Delta]$ that are still block diagonal in
$\omega_n$ but no longer in $\mathbf{k}$. The $2 \times 2$ blocks of
the correction matrices $\delta G^{-1} [\Delta]$ and $\delta G
[\Delta]$ are
\begin{eqnarray}
\delta G_{\mathbf{k}, \mathbf{k}', \omega_n}^{-1} [\Delta] &\equiv&
\delta G_{(\mathbf{k}, \omega_n), (\mathbf{k}', \omega_n)}^{-1}
[\Delta] = \left( \begin{array}{cc} -\delta \hat{\mu}_{\mathbf{k} -
\mathbf{k}'} & i p_{\mathbf{k}, \mathbf{k}'}^{\phantom{*}} \delta
\Delta_{\mathbf{k} - \mathbf{k}'}^{+} + i p_{\mathbf{k}',
\mathbf{k}}^{*} \delta \Delta_{\mathbf{k} - \mathbf{k}'}^{-} \\ -i
p_{\mathbf{k}', \mathbf{k}}^{*} \big( \delta \Delta_{\mathbf{k}' -
\mathbf{k}}^{+} \big)^{*} - i p_{\mathbf{k},
\mathbf{k}'}^{\phantom{*}} \big( \delta \Delta_{\mathbf{k}' -
\mathbf{k}}^{-} \big)^{*} & \delta \hat{\mu}_{\mathbf{k} -
\mathbf{k}'} \end{array} \right), \nonumber \\
\label{eq-corr-G} \\
\delta G_{\mathbf{k}, \mathbf{k}', \omega_n} [\Delta] &\equiv&
\delta G_{(\mathbf{k}, \omega_n), (\mathbf{k}', \omega_n)} [\Delta]
= -G_{\mathbf{k}, \omega_n} [\Delta] \cdot \delta G_{\mathbf{k},
\mathbf{k}', \omega_n}^{-1} [\Delta] \cdot G_{\mathbf{k}', \omega_n}
[\Delta]. \nonumber
\end{eqnarray}
Up to linear order in both $\delta \hat{\mu}_{\mathbf{q}}$ and
$\delta \Delta_{\mathbf{q}}^{\pm}$, the resulting correction to the
saddle-point equation in Eq.~(\ref{eq-mf-S-2}) is then
\begin{eqnarray}
\delta \Delta_{\mathbf{q}}^{\pm} &=& \frac{g} {2N \beta}
\sum_{\mathbf{k}', \mathbf{k}'', \omega_n} \mathrm{Tr} \left\{
\delta G_{\mathbf{k}', \mathbf{k}'', \omega_n} [\Delta] \cdot
\frac{\partial (\delta G_{\mathbf{k}'', \mathbf{k}', \omega_n}^{-1}
[\Delta])} {\partial (\delta \Delta_{\mathbf{q}}^{\pm})^{*}}
\right\} \label{eq-corr-S-1} \\
&=& -\frac{g} {2N \beta} \sum_{\mathbf{k}, \omega_n} \mathrm{Tr}
\left\{ G_{\mathbf{k} + \frac{1}{2} \mathbf{q}, \omega_n} [\Delta]
\cdot \delta G_{\mathbf{k} + \frac{1}{2} \mathbf{q}, \mathbf{k} -
\frac{1}{2} \mathbf{q}, \omega_n}^{-1} [\Delta] \cdot G_{\mathbf{k}
- \frac{1}{2} \mathbf{q}, \omega_n} [\Delta] \cdot \frac{\partial
(\delta G_{\mathbf{k} - \frac{1}{2} \mathbf{q}, \mathbf{k} +
\frac{1}{2} \mathbf{q}, \omega_n}^{-1} [\Delta])} {\partial (\delta
\Delta_{\mathbf{q}}^{\pm})^{*}} \right\}. \nonumber
\end{eqnarray}
Substituting Eqs.~(\ref{eq-mf-G}) and (\ref{eq-corr-G}) into
Eq.~(\ref{eq-corr-S-1}), the saddle-point equations for $\delta
\Delta_{\mathbf{q}}^{+}$ and $\delta \Delta_{\mathbf{q}}^{-}$ can be
written as
\begin{eqnarray}
\delta \Delta_{\mathbf{q}}^{+} &=& A_{\mathbf{q}}^{+} \delta
\hat{\mu}_{\mathbf{q}}^{\phantom{*}} + \left( 1 -
B_{\mathbf{q}}^{++} \right) \delta \Delta_{\mathbf{q}}^{+} -
B_{\mathbf{q}}^{+-} \delta \Delta_{\mathbf{q}}^{-} -
C_{\mathbf{q}}^{++} (\delta \Delta_{-\mathbf{q}}^{+})^{*} -
C_{\mathbf{q}}^{+-} (\delta
\Delta_{-\mathbf{q}}^{-})^{*}, \nonumber \\
\delta \Delta_{\mathbf{q}}^{-} &=& A_{\mathbf{q}}^{-} \delta
\hat{\mu}_{\mathbf{q}}^{\phantom{*}} - B_{\mathbf{q}}^{-+} \delta
\Delta_{\mathbf{q}}^{+} + \left( 1 - B_{\mathbf{q}}^{--} \right)
\delta \Delta_{\mathbf{q}}^{-} - C_{\mathbf{q}}^{-+} (\delta
\Delta_{-\mathbf{q}}^{+})^{*} - C_{\mathbf{q}}^{--} (\delta
\Delta_{-\mathbf{q}}^{-})^{*}. \label{eq-corr-S-2}
\end{eqnarray}
For $|\mathbf{q}| \ll |\mathbf{k}| \sim k_F$, where $|p_{\mathbf{k}
\pm \mathbf{q}'} - p_{\mathbf{k}}| \ll |p_{\mathbf{k}}|$ and thus
$p_{\mathbf{k} \pm \mathbf{q}'} \approx p_{\mathbf{k}}$ in terms of
$\mathbf{q}' \equiv \mathbf{q}/2$, the coefficients in
Eq.~(\ref{eq-corr-S-2}) are
\begin{eqnarray}
&& A_{\mathbf{q}}^{+} = \frac{g} {2N \beta} \sum_{\mathbf{k},
\omega_n} \frac{\bar{\Delta} |p_{\mathbf{k}}^{\phantom{*}}|^2 \big(
\varepsilon_{\mathbf{k} + \mathbf{q}'}^{\phantom{*}} +
\varepsilon_{\mathbf{k} - \mathbf{q}'}^{\phantom{*}} \big)} {\big(
\omega_n^2 + \varepsilon_{\mathbf{k} + \mathbf{q}'}^2 +
\bar{\Delta}^2 |p_{\mathbf{k}}^{\phantom{*}}|^2 \big) \big(
\omega_n^2 + \varepsilon_{\mathbf{k} - \mathbf{q}'}^2 +
\bar{\Delta}^2 |p_{\mathbf{k}}^{\phantom{*}}|^2
\big)}, \nonumber \\
&& A_{\mathbf{q}}^{-} = \frac{g} {2N \beta} \sum_{\mathbf{k},
\omega_n} \frac{\bar{\Delta} p_{\mathbf{k}}^2 \big(
\varepsilon_{\mathbf{k} + \mathbf{q}'}^{\phantom{*}} +
\varepsilon_{\mathbf{k} - \mathbf{q}'}^{\phantom{*}} \big)} {\big(
\omega_n^2 + \varepsilon_{\mathbf{k} + \mathbf{q}'}^2 +
\bar{\Delta}^2 |p_{\mathbf{k}}^{\phantom{*}}|^2 \big) \big(
\omega_n^2 + \varepsilon_{\mathbf{k} - \mathbf{q}'}^2 +
\bar{\Delta}^2 |p_{\mathbf{k}}^{\phantom{*}}|^2
\big)}, \nonumber \\
&& B_{\mathbf{q}}^{++} = B_{\mathbf{q}}^{--} = 1 - \frac{g} {2N
\beta} \sum_{\mathbf{k}, \omega_n}
\frac{|p_{\mathbf{k}}^{\phantom{*}}|^2 \big( \varepsilon_{\mathbf{k}
+ \mathbf{q}'}^{\phantom{*}} \varepsilon_{\mathbf{k} -
\mathbf{q}'}^{\phantom{*}} + \omega_n^2 \big)} {\big( \omega_n^2 +
\varepsilon_{\mathbf{k} + \mathbf{q}'}^2 + \bar{\Delta}^2
|p_{\mathbf{k}}^{\phantom{*}}|^2 \big) \big( \omega_n^2 +
\varepsilon_{\mathbf{k} - \mathbf{q}'}^2 + \bar{\Delta}^2
|p_{\mathbf{k}}^{\phantom{*}}|^2
\big)} \nonumber \\
&& \qquad \,\, = \Bigg\{ 1 - \frac{g} {4N \beta} \sum_{\mathbf{k},
\omega_n} \sum_{\pm} \frac{|p_{\mathbf{k}}^{\phantom{*}}|^2}
{\omega_n^2 + \varepsilon_{\mathbf{k} \pm \mathbf{q}'}^2 +
\bar{\Delta}^2 |p_{\mathbf{k}}^{\phantom{*}}|^2} \Bigg\} + \frac{g}
{4N \beta} \sum_{\mathbf{k}, \omega_n}
\frac{|p_{\mathbf{k}}^{\phantom{*}}|^2 \big[ \big(
\varepsilon_{\mathbf{k} + \mathbf{q}'}^{\phantom{*}} -
\varepsilon_{\mathbf{k} - \mathbf{q}'}^{\phantom{*}} \big)^2 + 2
\bar{\Delta}^2 |p_{\mathbf{k}}^{\phantom{*}}|^2 \big]} {\big(
\omega_n^2 + \varepsilon_{\mathbf{k} + \mathbf{q}'}^2 +
\bar{\Delta}^2 |p_{\mathbf{k}}^{\phantom{*}}|^2 \big) \big(
\omega_n^2 + \varepsilon_{\mathbf{k} - \mathbf{q}'}^2 +
\bar{\Delta}^2 |p_{\mathbf{k}}^{\phantom{*}}|^2
\big)}, \nonumber \\
&& B_{\mathbf{q}}^{-+} = (B_{\mathbf{q}}^{+-})^{*} = - \frac{g} {2N
\beta} \sum_{\mathbf{k}, \omega_n} \frac{p_{\mathbf{k}}^2 \big(
\varepsilon_{\mathbf{k} + \mathbf{q}'}^{\phantom{*}}
\varepsilon_{\mathbf{k} - \mathbf{q}'}^{\phantom{*}} + \omega_n^2
\big)} {\big( \omega_n^2 + \varepsilon_{\mathbf{k} + \mathbf{q}'}^2
+ \bar{\Delta}^2 |p_{\mathbf{k}}^{\phantom{*}}|^2 \big) \big(
\omega_n^2 + \varepsilon_{\mathbf{k} - \mathbf{q}'}^2 +
\bar{\Delta}^2 |p_{\mathbf{k}}^{\phantom{*}}|^2
\big)} \label{eq-corr-ABC} \\
&& \qquad \,\, = -\Bigg\{ \frac{g} {4N \beta} \sum_{\mathbf{k},
\omega_n} \sum_{\pm} \frac{p_{\mathbf{k}}^2} {\omega_n^2 +
\varepsilon_{\mathbf{k} \pm \mathbf{q}'}^2 + \bar{\Delta}^2
|p_{\mathbf{k}}^{\phantom{*}}|^2} \Bigg\} + \frac{g} {4N \beta}
\sum_{\mathbf{k}, \omega_n} \frac{p_{\mathbf{k}}^2 \big[ \big(
\varepsilon_{\mathbf{k} + \mathbf{q}'}^{\phantom{*}} -
\varepsilon_{\mathbf{k} - \mathbf{q}'}^{\phantom{*}} \big)^2 + 2
\bar{\Delta}^2 |p_{\mathbf{k}}^{\phantom{*}}|^2 \big]} {\big(
\omega_n^2 + \varepsilon_{\mathbf{k} + \mathbf{q}'}^2 +
\bar{\Delta}^2 |p_{\mathbf{k}}^{\phantom{*}}|^2 \big) \big(
\omega_n^2 + \varepsilon_{\mathbf{k} - \mathbf{q}'}^2 +
\bar{\Delta}^2 |p_{\mathbf{k}}^{\phantom{*}}|^2
\big)}, \nonumber \\
&& C_{\mathbf{q}}^{++} = \frac{g} {2N \beta} \sum_{\mathbf{k},
\omega_n} \frac{\bar{\Delta}^2 |p_{\mathbf{k}}^{\phantom{*}}|^4}
{\big( \omega_n^2 + \varepsilon_{\mathbf{k} + \mathbf{q}'}^2 +
\bar{\Delta}^2 |p_{\mathbf{k}}^{\phantom{*}}|^2 \big) \big(
\omega_n^2 + \varepsilon_{\mathbf{k} - \mathbf{q}'}^2 +
\bar{\Delta}^2 |p_{\mathbf{k}}^{\phantom{*}}|^2
\big)}, \nonumber \\
&& C_{\mathbf{q}}^{+-} = C_{\mathbf{q}}^{-+} = \frac{g} {2N \beta}
\sum_{\mathbf{k}, \omega_n} \frac{\bar{\Delta}^2 p_{\mathbf{k}}^2 \,
|p_{\mathbf{k}}^{\phantom{*}}|^2} {\big( \omega_n^2 +
\varepsilon_{\mathbf{k} + \mathbf{q}'}^2 + \bar{\Delta}^2
|p_{\mathbf{k}}^{\phantom{*}}|^2 \big) \big( \omega_n^2 +
\varepsilon_{\mathbf{k} - \mathbf{q}'}^2 + \bar{\Delta}^2
|p_{\mathbf{k}}^{\phantom{*}}|^2
\big)}, \nonumber \\
&& C_{\mathbf{q}}^{--} = \frac{g} {2N \beta} \sum_{\mathbf{k},
\omega_n} \frac{\bar{\Delta}^2 p_{\mathbf{k}}^4} {\big( \omega_n^2 +
\varepsilon_{\mathbf{k} + \mathbf{q}'}^2 + \bar{\Delta}^2
|p_{\mathbf{k}}^{\phantom{*}}|^2 \big) \big( \omega_n^2 +
\varepsilon_{\mathbf{k} - \mathbf{q}'}^2 + \bar{\Delta}^2
|p_{\mathbf{k}}^{\phantom{*}}|^2 \big)}, \nonumber
\end{eqnarray}
where we assume without loss of generality that $\bar{\Delta}$ is
real and positive. Using $p_{\mathbf{k} \pm \mathbf{q}'} \approx
p_{\mathbf{k}}$, the terms in the curly brackets vanish for
$B_{\mathbf{q}}^{++}$ because of Eq.~(\ref{eq-mf-S-2}) and for
$B_{\mathbf{q}}^{-+}$ because $p_{\mathbf{k}}^2$ changes sign under
fourfold rotation symmetry. Since $\varepsilon_{\mathbf{k} +
\mathbf{q}'} + \varepsilon_{\mathbf{k} - \mathbf{q}'} \approx 2
\varepsilon_{\mathbf{k}}$ for $|\mathbf{q}| \ll |\mathbf{k}|$, the
summands of $A_{\mathbf{q}}^{\pm}$ change sign at the Fermi surface,
$\varepsilon_{\mathbf{k}} = 0$, and the dominant contributions to
the resulting sums are from regions far away from the Fermi surface.
Given that $|\varepsilon_{\mathbf{k} \pm \mathbf{q}'} -
\varepsilon_{\mathbf{k}}| \ll |\varepsilon_{\mathbf{k}}|$ in those
regions, it is then reasonable to approximate $A_{\mathbf{q}}^{\pm}$
with $A_{\mathbf{0}}^{\pm}$ for $|\mathbf{q}| \ll k_F$. In this
approximation, we obtain
\begin{eqnarray}
&& A_{\mathbf{q}}^{+} \approx A_{\mathbf{0}}^{+} = \frac{g} {2N
\beta} \sum_{\mathbf{k}, \omega_n} \frac{2 \bar{\Delta}
|p_{\mathbf{k}}^{\phantom{*}}|^2
\varepsilon_{\mathbf{k}}^{\phantom{*}}} {\big( \omega_n^2 +
\varepsilon_{\mathbf{k}}^2 + \bar{\Delta}^2
|p_{\mathbf{k}}^{\phantom{*}}|^2 \big)^2} = \bar{\Delta} \,
\frac{\partial} {\partial \varepsilon_F} \Bigg[ \frac{g} {2N \beta}
\sum_{\mathbf{k}, \omega_n} \frac{|p_{\mathbf{k}}^{\phantom{*}}|^2}
{\omega_n^2 + \varepsilon_{\mathbf{k}}^2 + \bar{\Delta}^2
|p_{\mathbf{k}}^{\phantom{*}}|^2} \Bigg], \nonumber \\
&& A_{\mathbf{q}}^{-} \approx A_{\mathbf{0}}^{-} = \frac{g} {2N
\beta} \sum_{\mathbf{k}, \omega_n} \frac{2 \bar{\Delta}
p_{\mathbf{k}}^2 \varepsilon_{\mathbf{k}}^{\phantom{*}}} {\big(
\omega_n^2 + \varepsilon_{\mathbf{k}}^2 + \bar{\Delta}^2
|p_{\mathbf{k}}^{\phantom{*}}|^2 \big)^2} = 0, \label{eq-corr-A}
\end{eqnarray}
where $A_{\mathbf{0}}^{-}$ vanishes because $p_{\mathbf{k}}^2$
changes sign under fourfold rotation symmetry. In contrast, the
dominant contributions to the sums of $B_{\mathbf{q}}^{\pm \pm}$ and
$C_{\mathbf{q}}^{\pm \pm}$ are from the vicinity of the Fermi
surface. Therefore, we can write $p_{\mathbf{k}} \approx 2 k_F e^{i
\vartheta_{\mathbf{k}}} \equiv p_F e^{i \vartheta_{\mathbf{k}}}$ and
\begin{equation}
\varepsilon_{\mathbf{k} \pm \mathbf{q}'} \approx
\varepsilon_{\mathbf{k}} \pm \cos (\vartheta_{\mathbf{k}} -
\vartheta_{\mathbf{q}'}) \, v_F |\mathbf{q}'| =
\varepsilon_{\mathbf{k}} \pm \frac{1}{2} \cos
(\vartheta_{\mathbf{k}} - \vartheta_{\mathbf{q}}) \, v_F
|\mathbf{q}| = \varepsilon_{\mathbf{k}} \pm \cos
(\vartheta_{\mathbf{k}} - \vartheta_{\mathbf{q}}) \, \bar{\Delta}
p_F \, \frac{\xi |\mathbf{q}|} {2}, \label{eq-corr-epsilon}
\end{equation}
where $v_F = k_F / m$ is the Fermi velocity, $\xi = v_F / (2k_F
\bar{\Delta}) = 1 / (2m \bar{\Delta})$ is the superconducting
coherence length, while $\vartheta_{\mathbf{k}}$ is the angle
between $\mathbf{k}$ and $\hat{\mathbf{r}}_x$. If we then take the
zero-temperature limit ($\beta \rightarrow \infty$), and turn the
summation in $\mathbf{k}$ into an integral,
\begin{equation}
\frac{1}{N} \sum_{\mathbf{k}} \rightarrow \frac{\nu} {2\pi}
\int_{-\infty}^{+\infty} d\varepsilon \int_{0}^{2\pi} d\vartheta,
\label{eq-corr-k}
\end{equation}
where $\varepsilon = \varepsilon_{\mathbf{k}}$ and $\vartheta =
\vartheta_{\mathbf{k}} - \vartheta_{\mathbf{q}}$, while $\nu$ is the
density of states at the Fermi level, the coefficients
$B_{\mathbf{q}}^{\pm \pm}$ and $C_{\mathbf{q}}^{\pm \pm}$ become
\begin{eqnarray}
&& B_{\mathbf{q}}^{++} = B_{\mathbf{q}}^{--} = B_{\mathbf{0}}^{++}
\left[ I_0 \left( \frac{\xi |\mathbf{q}|} {2} \right) + I_0' \left(
\frac{\xi |\mathbf{q}|} {2} \right) \right],
\nonumber \\
&& B_{\mathbf{q}}^{-+} = (B_{\mathbf{q}}^{+-})^{*} =
B_{\mathbf{0}}^{++} e^{2i \vartheta_{\mathbf{q}}} \left[ I_2 \left(
\frac{\xi |\mathbf{q}|} {2} \right) + I_2' \left( \frac{\xi
|\mathbf{q}|} {2} \right) \right],
\nonumber \\
&& C_{\mathbf{q}}^{++} = B_{\mathbf{0}}^{++} \, I_0 \left( \frac{\xi
|\mathbf{q}|} {2} \right), \label{eq-corr-BC} \\
&& C_{\mathbf{q}}^{+-} = C_{\mathbf{q}}^{-+} = B_{\mathbf{0}}^{++}
e^{2i \vartheta_{\mathbf{q}}} \, I_2 \left( \frac{\xi |\mathbf{q}|}
{2} \right), \nonumber \\
&& C_{\mathbf{q}}^{--} = B_{\mathbf{0}}^{++} e^{4i
\vartheta_{\mathbf{q}}} \, I_4 \left( \frac{\xi |\mathbf{q}|} {2}
\right), \nonumber
\end{eqnarray}
where the common constant of proportionality is given by
\begin{equation}
B_{\mathbf{0}}^{++} = \frac{g} {2N \beta} \sum_{\mathbf{k},
\omega_n} \frac{\bar{\Delta}^2 |p_{\mathbf{k}}^{\phantom{*}}|^4}
{\big( \omega_n^2 + \varepsilon_{\mathbf{k}}^2 + \bar{\Delta}^2
|p_{\mathbf{k}}^{\phantom{*}}|^2 \big)^2} = -\frac{1}{2}
\bar{\Delta} \, \frac{\partial} {\partial \bar{\Delta}} \Bigg[
\frac{g} {2N \beta} \sum_{\mathbf{k}, \omega_n}
\frac{|p_{\mathbf{k}}^{\phantom{*}}|^2} {\omega_n^2 +
\varepsilon_{\mathbf{k}}^2 + \bar{\Delta}^2
|p_{\mathbf{k}}^{\phantom{*}}|^2} \Bigg], \label{eq-corr-B}
\end{equation}
while the dimensionless functions are appropriate integrals,
\begin{eqnarray}
I_n (x) &=& \frac{1} {2\pi^2} \int_{-\infty}^{+\infty}
d\tilde{\omega} \int_{-\infty}^{+\infty} d\tilde{\varepsilon}
\int_{0}^{2\pi} d\vartheta \, \frac{e^{i n \vartheta}} {\big[ 1 +
\tilde{\omega}^2 + (\tilde{\varepsilon} + x \cos \vartheta)^2 \big]
\big[ 1 + \tilde{\omega}^2 + (\tilde{\varepsilon} - x \cos
\vartheta)^2
\big]}, \nonumber \\
I_n' (x) &=& \frac{1} {2\pi^2} \int_{-\infty}^{+\infty}
d\tilde{\omega} \int_{-\infty}^{+\infty} d\tilde{\varepsilon}
\int_{0}^{2\pi} d\vartheta \, \frac{2x^2 \cos^2 \vartheta \, e^{i n
\vartheta}} {\big[ 1 + \tilde{\omega}^2 + (\tilde{\varepsilon} + x
\cos \vartheta)^2 \big] \big[ 1 + \tilde{\omega}^2 +
(\tilde{\varepsilon} - x \cos \vartheta)^2 \big]},
\label{eq-corr-I-1}
\end{eqnarray}
with $\tilde{\omega} = \omega / (\bar{\Delta} p_F)$ and
$\tilde{\varepsilon} = \varepsilon / (\bar{\Delta} p_F)$, taking the
exact analytical forms
\begin{eqnarray}
&& I_0 (x) = \frac{\arctan x} {x}, \qquad I_2 (x) = \frac{\log (1 +
x^2) - x \arctan x} {x^2}, \qquad I_4 (x) = \frac{x^2 (2 + x \arctan
x) - 2(1 + x^2) \log (1 + x^2)} {x^4}, \nonumber \\
&& I_0' (x) = \log (1 + x^2), \qquad I_2' (x) = 1 - \frac{\log (1 +
x^2)} {x^2}. \label{eq-corr-I-2}
\end{eqnarray}
Using Eqs.~(\ref{eq-corr-A}) and (\ref{eq-corr-BC}), the solution of
Eq.~(\ref{eq-corr-S-2}) for $\delta \Delta_{\mathbf{q}}^{\pm}$ then
becomes
\begin{equation}
\delta \Delta_{\mathbf{q}}^{+} = f \left( \frac{\xi |\mathbf{q}|}
{2} \right) \frac{A_{\mathbf{0}}^{+}} {2 B_{\mathbf{0}}^{++}} \,
\delta \hat{\mu}_{\mathbf{q}}, \qquad \delta \Delta_{\mathbf{q}}^{-}
= -h \left( \frac{\xi |\mathbf{q}|} {2} \right) e^{2i
\vartheta_{\mathbf{q}}} \, \frac{A_{\mathbf{0}}^{+}} {2
B_{\mathbf{0}}^{++}} \, \delta \hat{\mu}_{\mathbf{q}},
\label{eq-corr-delta-2}
\end{equation}
where the dimensionless functions,
\begin{eqnarray}
f(x) &=& \frac{2I_0 (x) + 2I_4 (x) + 2I_0' (x)} {[2I_0 (x) + I_0'
(x)] [I_0 (x) + I_4 (x) + I_0' (x)] - [2I_2 (x) + I_2' (x)]^2},
\nonumber
\\
h(x) &=& \frac{4I_2 (x) + 2I_2' (x)} {[2I_0 (x) + I_0' (x)] [I_0 (x)
+ I_4 (x) + I_0' (x)] - [2I_2 (x) + I_2' (x)]^2},
\label{eq-corr-fh-1}
\end{eqnarray}
are plotted in Fig.~\ref{fig-S} and have asymptotic forms
\begin{equation}
f(x) \approx \bigg{\{} \begin{array}{c} 1 - \frac{x^2}{6} \quad (x
\ll 1), \\ \frac{1} {\ln x} \qquad \,\, (x \gg 1),
\end{array} \qquad h(x) \approx \bigg{\{} \begin{array}{c} \frac{x^2}{6}
\qquad \,\,\,\, (x \ll 1), \\ \frac{1} {2 (\ln x)^2} \,\,\,\,\, (x
\gg 1). \end{array} \label{eq-corr-fh-2}
\end{equation}
Finally, if we rewrite the disorder-free saddle-point equation in
Eq.~(\ref{eq-mf-S-2}) as
\begin{equation}
P [\varepsilon_F, \bar{\Delta} (\varepsilon_F)] \equiv \frac{g} {2N
\beta} \sum_{\mathbf{k}, \omega_n}
\frac{|p_{\mathbf{k}}^{\phantom{*}}|^2} {\omega_n^2 +
\varepsilon_{\mathbf{k}}^2 + \bar{\Delta}^2
|p_{\mathbf{k}}^{\phantom{*}}|^2} = 1, \label{eq-corr-P-1}
\end{equation}
take its total derivative with respect to $\varepsilon_F$, and
substitute $\partial P / \partial \varepsilon_F$ and $\partial P /
\partial \bar{\Delta}$ from Eqs.~(\ref{eq-corr-A}) and
(\ref{eq-corr-B}), we obtain
\begin{equation}
\frac{dP} {d\varepsilon_F} = \frac{\partial P} {\partial
\varepsilon_F} + \frac{\partial P} {\partial \bar{\Delta}}
\frac{\partial \bar{\Delta}} {\partial \varepsilon_F} = \frac{1}
{\bar{\Delta}} \left( A_{\mathbf{0}}^{+} - 2B_{\mathbf{0}}^{++}
\frac{\partial \bar{\Delta}} {\partial \varepsilon_F} \right) = 0.
\label{eq-corr-P-2}
\end{equation}
Therefore, $A_{\mathbf{0}}^{+} / (2B_{\mathbf{0}}^{++}) = \partial
\bar{\Delta} / \partial \varepsilon_F$, and
Eq.~(\ref{eq-corr-delta-2}) is equivalent to Eq.~(10) in the main
text.

\begin{figure}[t]
\centering
\includegraphics[width=0.75\columnwidth]{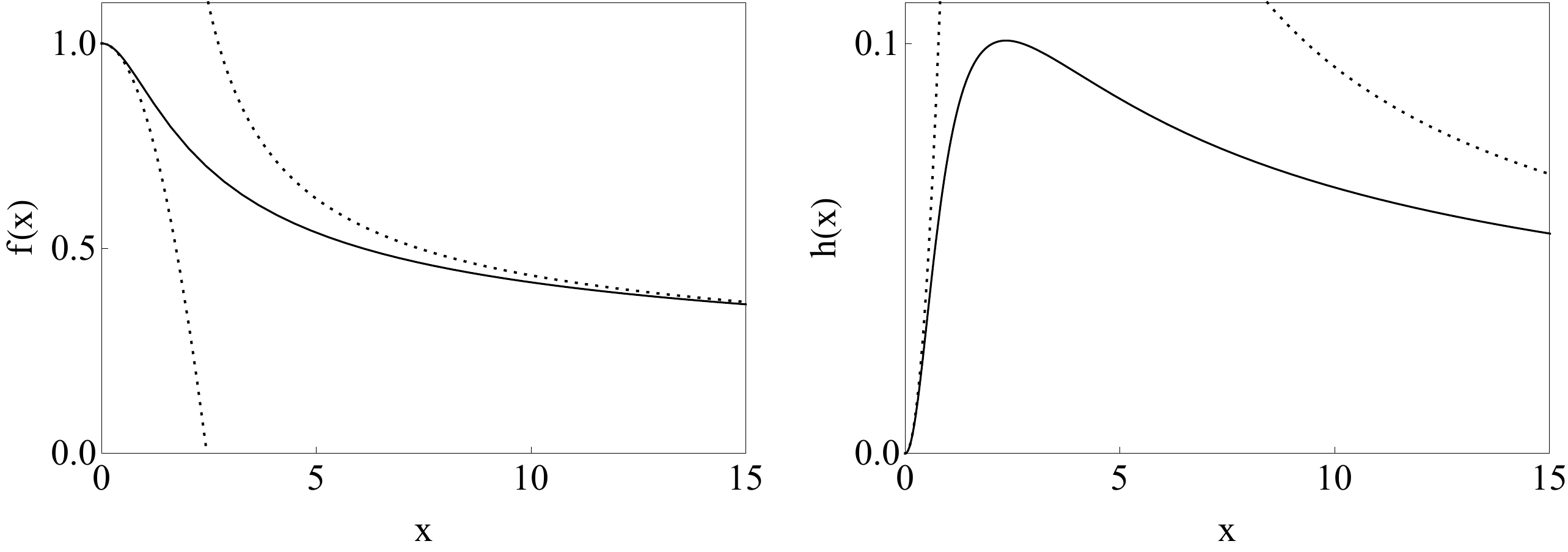}
\caption{Exact forms (solid lines) and asymptotic forms (dotted
lines) of the dimensionless functions $f(x)$ and $h(x)$.}
\label{fig-S}
\end{figure}

\section{Self-consistent numerical solution}

Here we describe the details of the numerical procedure that we use
to obtain the self-consistent solution of the Bogoliubov-de Gennes
(BdG) Hamiltonian [see Eq.~(4) in the main text],
\begin{equation}
H = -\sum_{\mathbf{r}} \mu_{\mathbf{r}}^{\phantom{\dag}}
c_{\mathbf{r}}^{\dag} c_{\mathbf{r}}^{\phantom{\dag}} -
\sum_{\langle \mathbf{r}, \mathbf{r}' \rangle} \left( t_{\mathbf{r},
\mathbf{r}'}^{\phantom{*}} c_{\mathbf{r}}^{\dag}
c_{\mathbf{r}'}^{\phantom{\dag}} + t_{\mathbf{r}, \mathbf{r}'}^{*}
c_{\mathbf{r}'}^{\dag} c_{\mathbf{r}}^{\phantom{\dag}} \right) -
\sum_{\langle \mathbf{r}, \mathbf{r}' \rangle} \left(
\Delta_{\mathbf{r}, \mathbf{r}'}^{*} c_{\mathbf{r}}^{\phantom{\dag}}
c_{\mathbf{r}'}^{\phantom{\dag}} + \Delta_{\mathbf{r},
\mathbf{r}'}^{\phantom{*}} c_{\mathbf{r}'}^{\dag}
c_{\mathbf{r}}^{\dag} \right), \label{eq-num-H}
\end{equation}
in terms of the superconducting pairing potentials [see Eq.~(5) in
the main text],
\begin{equation}
\Delta_{\mathbf{r}, \mathbf{r}'} = -\Delta_{\mathbf{r}', \mathbf{r}}
= g \langle c_{\mathbf{r}} c_{\mathbf{r}'} \rangle.
\label{eq-num-delta-1}
\end{equation}
The site-dependent chemical potentials are $\mu_{\mathbf{r}} =
\bar{\mu} + \delta \mu_{\mathbf{r}}$, while the hopping amplitudes
are $t_{\mathbf{r}, \mathbf{r}'} = t$ for a vortex-free system and
$t_{\mathbf{r}, \mathbf{r}'} = t u_{\mathbf{r}, \mathbf{r'}} e^{i
A'_{\mathbf{r}, \mathbf{r}'}}$ if there are two vortices at
positions $\mathbf{R}_{1,2}$ connected by a $\mathbb{Z}_2$ flux
string [see Fig.~1(a) in the main text], where $u_{\mathbf{r},
\mathbf{r'}}$ is $-1$ ($+1$) if the $\mathbb{Z}_2$ flux string
intersects (does not intersect) the link $\langle \mathbf{r},
\mathbf{r}' \rangle$, and
\begin{equation}
A'_{\mathbf{r}, \mathbf{r}'} = \int_{\mathbf{r}}^{\mathbf{r}'}
\left[ \mathbf{A}' (\hat{\mathbf{r}} - \mathbf{R}_1) + \mathbf{A}'
(\hat{\mathbf{r}} - \mathbf{R}_2) \right] \cdot d \hat{\mathbf{r}}
\label{eq-num-A-1}
\end{equation}
corresponds to an effective vector potential which is only nonzero
within a London penetration depth $\lambda$ of each vortex. Indeed,
the components of $\mathbf{A}' (\mathbf{r})$ in polar coordinates,
$\mathbf{r} = (r, \vartheta)$, can be written as
\begin{equation}
A_r' (\mathbf{r}) = 0, \qquad A_{\vartheta}' (\mathbf{r}) =
-\frac{a(r)} {2r}, \label{eq-num-A-2}
\end{equation}
where $a(r)$ must asymptotically satisfy $a(r) \approx 1$ for $r \ll
\lambda$ and $a(r) \sim e^{-r / \lambda}$ for $r \gg \lambda$. We
choose $a(r) = (1 + r / \lambda) \, e^{-r / \lambda}$ but note that
the precise form of $a(r)$ does not matter as long as the asymptotic
conditions are satisfied.

To find a self-consistent solution of Eqs.~(\ref{eq-num-H}) and
(\ref{eq-num-delta-1}), we first make an initial guess for the
pairing potentials $\Delta_{\mathbf{r}, \mathbf{r}'}$. Then, we
solve the BdG Hamiltonian in Eq.~(\ref{eq-num-H}) by substituting
these pairing potentials, and compute the ground-state expectation
values in Eq.~(\ref{eq-num-delta-1}) to obtain an updated set of
pairing potentials. Finally, we repeat this procedure iteratively
until the pairing potentials converge up to the desired accuracy. In
practice, we always start from a vortex-free system in the
disorder-free limit. To describe a $p_x + i p_y$ superconductor with
the correct pairing symmetry, the appropriate initial guess is
\begin{equation}
\Delta_{\mathbf{r}, \mathbf{r} + \hat{\mathbf{r}}_x} = \Delta_0,
\qquad \Delta_{\mathbf{r}, \mathbf{r} + \hat{\mathbf{r}}_y} = i
\Delta_0. \label{eq-num-delta-2}
\end{equation}
Due to the symmetries of the system, the iterative procedure does
not change the form of Eq.~(\ref{eq-num-delta-2}) but only makes
$\Delta_0$ converge to the right value $\bar{\Delta}$. The next step
is to introduce two vortices at positions $\mathbf{R}_{1,2}$ with an
appropriate initial guess,
\begin{eqnarray}
\Delta_{\mathbf{r}, \mathbf{r} + \hat{\mathbf{r}}_x} &=&
u_{\mathbf{r}, \mathbf{r} + \hat{\mathbf{r}}_x} \bar{\Delta} \tanh
\left[ \frac{1} {\xi} \left( \mathbf{r} + \frac{1}{2}
\hat{\mathbf{r}}_x - \mathbf{R}_1 \right) \right] \tanh \left[
\frac{1} {\xi} \left( \mathbf{r} + \frac{1}{2}
\hat{\mathbf{r}}_x - \mathbf{R}_2 \right) \right], \nonumber \\
\Delta_{\mathbf{r}, \mathbf{r} + \hat{\mathbf{r}}_y} &=& i
u_{\mathbf{r}, \mathbf{r} + \hat{\mathbf{r}}_y} \bar{\Delta} \tanh
\left[ \frac{1} {\xi} \left( \mathbf{r} + \frac{1}{2}
\hat{\mathbf{r}}_y - \mathbf{R}_1 \right) \right] \tanh \left[
\frac{1} {\xi} \left( \mathbf{r} + \frac{1}{2} \hat{\mathbf{r}}_y -
\mathbf{R}_2 \right) \right], \label{eq-num-delta-3}
\end{eqnarray}
where $\xi = 1 / (2m \bar{\Delta})$ is the superconducting coherence
length. Once the iterative procedure is converged, we use the
converged set of pairing potentials as the initial guess when we
finally introduce disorder. For each disorder realization, we turn
on disorder smoothly by fixing the given disorder realization and
only rescaling it (i.e., increasing its overall strength $\delta
\bar{\mu}$) in small steps. At each step, the initial guess for the
pairing potentials is the converged set of pairing potentials from
the previous step.

\clearpage

\end{widetext}



\begin{references}
\bibitem{Footnote} We also remark that our results
are not sensitive to $\lambda$ and even apply in the limit of
$\lambda \rightarrow 0$ or, equivalently, $A'_{\mathbf{r},
\mathbf{r}'} \rightarrow 0$.
\end{references}
\end{document}